\documentclass[journal ]{new-aiaa}
\usepackage[utf8]{inputenc}
\usepackage{textcomp}
\usepackage{float}
\usepackage{graphicx}
\usepackage{amsmath}
\usepackage[version=4]{mhchem}
\usepackage{siunitx}
\usepackage{bm}
\usepackage{longtable,tabularx}
\title{Data Mining-Based Escape-Family Analysis in The Multi-Body Models}

\author{Shuyue Fu \footnote{PhD Candidate, Shen Yuan Honors College, School of Astronautics, fushuyue@buaa.edu.cn.}, Di Wu \footnote{Associate Professor, School of Astronautics, wudi2025@buaa.edu.cn, Member AIAA (Corresponding Author).}, Shengping Gong \footnote{Professor, School of Astronautics, gongsp@buaa.edu.cn, Senior Member AIAA.}, and Peng Shi\footnote{Professor, School of Astronautics, shipeng@buaa.edu.cn.}}
\affil{Beihang University, Beijing, 100191, People's Republic of China}
\affil{State Key Laboratory of High-Efficiency Reusable Aerospace Transportation Technology, Beijing, 102206, People's Republic of China}

\begin{document}

\maketitle

\begin{abstract}
Escape trajectories from the Earth-Moon system play an important role in interplanetary transfer. This paper focuses on the escape trajectories from a 167 km circular Earth orbit in the Earth-Moon planar circular three-body problem and the Sun-Earth/Moon planar bicircular four-body problem and is denoted to providing a systematic analysis on these escape trajectories. To achieve these purposes, the solution space of escape trajectories are constructed, and escape trajectories with one lunar gravity assist are pre-filtered. Then, an effective method to identification escape families is proposed based on dynamical analysis and data mining techniques. Once the escape families are identified, the corresponding characteristics are analyzed to provide insights into the construction of escape trajectories. Based on these escape families, the effects of the solar gravity perturbation on the number of escape trajectories, the emergence and disappearance of escape families, variation in generalized energy, and transfer characteristics are further summarized, providing insights into the model selection in the escape trajectory construction. This paper establishes an analysis methodology of escape trajectories from a perspective of escape families, deepening the understanding of escape dynamics.
\end{abstract}

\section{Introduction}
\lettrine{R}{enewed} interest in interplanetary exploration has sparked as the proposal and implementation of several missions (e.g., \textit{Tianwen-1} \cite{yu2023tianwen} and \textit{Tianwen-3} \cite{hou2025search}), continuing human’s exploration of deep space. To achieve interplanetary transfer, the first step is to achieve the escape from the Earth-Moon system. To study escape trajectories from the Earth-Moon system, several scholars have proposed identification and construction methods of these trajectories. Zotos \cite{zotos2015classifying} proposed a distance criterion to identify escape trajectories in the Earth-Moon planar circular restricted three-body problem (PCR3BP), and calculated the escape basins of natural escape trajectories under several fixed Jacobi energy levels. Oshima \cite{oshima2021capture}, Boudad et al. \cite{boudad2022departure} and Pasquale et al. \cite{pasquale2024systematic} investigated escape trajectories in the Sun-Earth/Moon planar restricted four-body problem (PBCR4BP) from specific multi-body periodic orbits. Our previous works \cite{FU2025,fu2025energy} established an escape criterion combined with distance and generalized energy in the Earth-Moon PCR3BP based on Zotos’s work \cite{zotos2015classifying}, and further explored the construction method of escape trajectories with the lunar gravity assist (LGA). However, although the escape criteria and the construction methods of escape trajectories have been developed, the solution space of escape trajectories from the specific orbits has not been systematically investigated. Identifying the families of escape trajectories from a specific orbit in the solution space and investigating the corresponding characteristics of each family remains an open problem. Such investigation can further provide valuable prior knowledge about the construction of escape trajectories, including the distribution of initial states and energy variations. Meanwhile, the selection of the dynamical models also affect the analysis results. There have been several models applied to the construction of escape trajectories, including the Earth-Moon PCR3BP \cite{zotos2015classifying,FU2025,fu2025energy} and the Sun-Earth/Moon PBCR4BP \cite{mcelrath2012using,casalino2020design,boudad2022departure}. However, the difference between escape trajectories in the Earth-Moon PCR3BP and the Sun-Earth/Moon PBCR4BP has not been systematically explored. For Earth-Moon transfer scenarios, the solar gravity perturbation introduced by the Sun-Earth/Moon PBCR4BP plays an important role in achieving lunar ballistic capture and consequently reducing the Moon insertion impulse \cite{belbruno1993sun,oshima2019low}. For escape trajectories from the Earth-Moon system, the segments of trajectories in the Earth-Moon exterior region can be affected by the solar gravity perturbation. McElrath et al. \cite{mcelrath2012using} pointed out that the solar gravity perturbation may affect the variation in trajectory energy. Nevertheless, the justification for the model selection in the construction of escape trajectories still requires further quantitative support, such as the effects of the solar gravity perturbation on the transfer characteristics and the distribution of initial states. To address these problems, the purposes of this paper can be summarized as follows:

\begin{enumerate}[label=(\arabic*)]
\item To explore the solution space of escape trajectories from a specific orbits and consequently provide insights into the escape trajectory construction in the multi-body models.
\item To summarize the effects of the solar gravity perturbation on escape trajectories and provide insights into the model selection in the construction of escape trajectories.
\end{enumerate}

Before the exploration of the solution space, escape trajectories should be constructed in these two multi-body problems. Due to the absence of the analytical solutions in the Earth-Moon PCR3BP and the Sun-Earth/Moon PBCR4BP, escape trajectories should be constructed numerically. Similar to the construction of Earth-Moon transfers in the multi-body models, the construction methods of escape trajectories can be generally categorized into methods based on dynamical structures (e.g., reachable region near L2 libration point \cite{villac2003escaping,FHLX202506007} and energy transition domain \cite{fu2025energy}) and methods based on grid search \cite{FU2025}. The first method typically per-filters the initial states based on the use of dynamical structures, while the second method exhaustively searches the space of construction parameters and identifies trajectories satisfying the corresponding criterion, therefore presenting the distributions of initial states to perform a systematic analysis. Combining the features of these two types of construction methods with the purposes of this paper, we use the method based on grid search to construct escape trajectories in the Earth-Moon PCR3BP and the Sun-Earth/Moon PBCR4BP. We particularly focus on escape trajectories from a 167 km circular Earth orbit \cite{fu2025energy}. The solution space of escape trajectories in the Earth-Moon PCR3BP and the Sun-Earth/Moon PBCR4BP are constructed. Then, the solution space is analyzed. Before the analysis, pre-filter of escape trajectories are performed by the number of LGA. We typically focus on the escape trajectories with one LGA. The initial states of escape trajectories exhibits a clustered distribution pattern, implying that there exists different escape families. An effective method is proposed to identify escape families using the data mining techniques. These techniques have been used to classify trajectories based on Poincaré maps \cite{bosanac2020data,stefano2021applying}, classify transit orbits \cite{fu2025four} and low-energy transfer trajectories \cite{campana2024clustering}, identify periodic orbits \cite{smith2022constructing}, and investigate the trajectory sets near the periodic orbits \cite{nakhjiri2015automated,spear2025clustering}. Typical data mining techniques involve \textit{k}-means method \cite{ahmed2020k} and density-based spatial clustering of applications with noise (DBSCAN) method \cite{ester1996density}. Among these, the DBSCAN method is utilized to identify escape families because it is able to identify clusters of arbitrary shapes and does not required to the number of clusters to be specified in advance \cite{raschka2022machine}. Once escape families in the Earth-Moon PCR3BP and the Sun-Earth/Moon PBCR4BP are identified, the corresponding characteristics are analyzed, and the comparison between escape families in the Earth-Moon PCR3BP and the Sun-Earth/Moon PBCR4BP reveals the effects of the solar gravity perturbation. These effects are summarized in terms of the number of escape trajectories, the emergence and disappearance of escape families, variation in generalized energy, and transfer characteristics. Above all, three main contributions of this paper can be summarized as follows:
\begin{enumerate}[label=(\arabic*)]
\item An effective identification method of escape families is proposed based on dynamical analysis and the DBSCAN method.
\item Based on the proposed method, we further analyze the characteristics of identified escape families, providing further insights into escape dynamics and escape trajectory construction.
\item By comparing the escape families in the Earth-Moon PCR3BP with those in the Sun-Earth/Moon PBCR4BP, we summarize the effects of the solar gravity perturbation, providing a useful insight into the model selection in the construction of escape trajectories. Based on the work of McElrath et al. \cite{mcelrath2012using}, this paper provides an analysis methodology of the effects of the solar gravity perturbation from a perspective of escape families.
\end{enumerate}

The rest of this paper is organized as follows. Section \ref{sec2} presents the dynamical models and escape criterion adopted in this paper. Section \ref{sec3} proposes the identification method of escape families based on dynamical analysis and the DBSCAN method. The identification results are presented and analyzed in Section \ref{sec4}, yielding the effects of the solar gravity perturbation. Finally, conclusions are drawn in Section \ref{sec5}.

\section{Escape Trajectories in The Multi-Body Models}\label{sec2}
This section presents the foundation of this paper, including the Earth-Moon PCR3BP, the Sun-Earth/Moon PBCR4BP, and the criterion to identify escape trajectories in these models.
\subsection{Earth-Moon PCR3BP}\label{subsec2.1}
The Earth-Moon PCR3BP \cite{mccarthy2021leveraging} is firstly introduced. Compared to the patched two-body model \cite{li2025trajectory}, this model provides a higher fidelity \cite{mccarthy2021leveraging}. In this model, the Earth, the Moon, and the spacecraft are assumed to move in the same plane, and the orbits of the Earth and the Moon around their barycenter are circular orbits. The Earth-Moon rotating frame \cite{FU20254993} is adopted to express the equations of motion (EOM). For further simplification, dimensionless units are set as follows: the mass unit (MU) is set as the combined mass of the Earth and the Moon ($m_{\text{Earth}}+m_{\text{Moon}}$), the length unit (LU) is set as the distance between the centers of the Earth and the Moon, and the time units (TU) is set as $T_{\text{EM}}/2\pi$ ($T_{\text{EM}}$ denotes the orbital period of the Earth and the Moon around their barycenter). The EOM of this model are expressed as:
\begin{equation}
\dot x =u \text{ }\text{ }\text{ }\dot y =v\text{ }\text{ }\text{ }\dot u ={2v + \frac{{\partial {\Omega _3}}}{{\partial x}}}\text{ }\text{ }\text{ }\dot v={ - 2u + \frac{{\partial {\Omega _3}}}{{\partial y}}}\label{eq1}
\end{equation}
\begin{equation}
{\Omega _3} = \frac{1}{2}\left( {{x^2} + {y^2} + \mu \left( {1 - \mu } \right)} \right) + \frac{{1 - \mu }}{{{r_1}}} + \frac{\mu }{{{r_2}}}\label{eq2}
\end{equation}
where $\bm{X}=\left[x,\text{ }y,\text{ }u,\text{ }v\right]^\text{T}$ denotes the state of the above EOM. The parameter $\mu=m_{\text{Moon}}/\left(m_{\text{Earth}}+m_{\text{Moon}}\right)$ denotes the mass parameter of the Earth-Moon PCR3BP, and $\Omega_3$ denotes the effective potential of the Earth-Moon PCR3BP. The distance between the spacecraft and the Earth ($r_1$), the distance between the spacecraft and the Moon ($r_2$), and the distance between the spacecraft and the Earth-Moon barycenter ($r$) are expressed as follows:
\begin{equation}
{r_1} = \sqrt {{{\left( {x + \mu } \right)}^2} + {y^2}} \text{ }\text{ }\text{ }\text{ }{r_2} = \sqrt {{{\left( {x + \mu - 1} \right)}^2} + {y^2}}\text{ }\text{ }\text{ }\text{ }{r} = \sqrt {{x^2} + {y^2}}\label{eq3}
\end{equation}
The specific values of the above parameters can be found in Ref. \cite{fu2025four}. We focus on trajectories escaping from the Earth-Moon systems, to further explore the effects of the solar gravity perturbation, trajectories in the Sun-Earth/Moon PBCR4BP (which can be considered as the Earth-Moon PCR3BP combined with the solar gravity perturbation) are also investigated. Subsequently, the model of the Sun-Earth/Moon PBCR4BP is presented.

\subsection{Sun-Earth/Moon PBCR4BP}\label{subsec2.3}
In this subsection, the EOM of the Sun-Earth/Moon PBCR4BP \cite{yin2023midcourse} is presented. In this model, the Sun, the Earth, the Moon, and the spacecraft are assumed to move in the same plane, where the Earth-Moon barycenter moves in a circular orbit around the Sun and the Earth/Moon moves in circular orbits around their barycenter. Adopting the same frame and dimensionless units mentioned in Section \ref{subsec2.1}, the EOM of the Sun-Earth/Moon PBCR4BP can be expressed as follows:
\begin{equation}
\dot x =u \text{ }\text{ }\text{ }\dot y =v\text{ }\text{ }\text{ }\dot u ={2v + \frac{{\partial {\Omega _4}}}{{\partial x}}}\text{ }\text{ }\text{ }\dot v={ - 2u + \frac{{\partial {\Omega _4}}}{{\partial y}}}\label{eq_PBCR4BP}
\end{equation}
\begin{equation}
{\Omega _4} = \frac{1}{2}\left( {{x^2} + {y^2} + \mu \left( {1 - \mu } \right)} \right) + \frac{{1 - \mu }}{{{r_1}}} + \frac{\mu }{{{r_2}}} + \frac{{{\mu_{\text{S}}}}}{{{r_3}}} - \frac{{{\mu_{\text{S}}}}}{{{\rho ^2}}}\left( {x\cos {\theta _{\text{S}}} + y\sin {\theta _{\text{S}}}} \right)\label{eq_omega4}
\end{equation}
where $\Omega_4$ denotes the effective potential of the Sun-Earth/Moon PBCR4BP, $\mu_{\text{S}}$ denotes the dimensionless mass of the Sun calculated by $\mu_{\text{S}}=m_{\text{Sun}}/\left(m_{\text{Earth}}+m_{\text{Moon}}\right)$, $\rho$ denotes the distance between the Sun and the Earth-Moon barycenter, and $\theta_\text{S}$ represents the solar phase angle calculated by $\theta_\text{S}\left(t\right)=\theta_{\text{S}i}+\omega_\text{S}t$ ($\omega_\text{S}$ is the solar angle velocity in the Earth-Moon rotating frame, $t+t_i$ is the current epoch, and $\theta_{\text{S}i}$ is the initial solar phase angle calculated by $\theta_{\text{S}i}=\omega_\text{S}t_i$). The distance between the spacecraft and the Sun ($r_3$) is expressed as follows:
\begin{equation}
{r_3} = \sqrt {{{\left( {x - \rho \cos {\theta _{\text{S}}}} \right)}^2} + {{\left( {y - \rho \sin {\theta _{\text{S}}}} \right)}^2}}\label{eq_r3}
\end{equation}
The values of the corresponding parameters can be found in Ref. \cite{fu2025four}. Since no analytical solutions exist in these two models, the variable step-size, variable order (VSVO) Adams-Bashforth-Moulton algorithm with absolute and relative tolerances set to $1 \times 10^{-13}$, is adopted to integrate trajectories numerically. 

\subsection{Escape Criterion}\label{subsec2.2}
In the Earth-Moon PCR3BP, escape trajectories (trajectoires escaping from the Earth-Moon system) can be defined as trajectories satisfying $r \to \infty$ when the integration time satisfying $t \to \infty$ \cite{zotos2015classifying}. Since this definition can not be achieved from a numerical perspective, numerical approximation of this definition should be adopted. Zotos \cite{zotos2015classifying} adopted the necessary condition of the above definition, i.e., when $r>R_d$ ($R_d$ denotes a preset distance threshold), trajectories can be identified as escape trajectories. Based on his work, our previous work \cite{FU2025} further developed an escape criterion based on $r$ and generalized energy $E$. This criterion can further exclude trajectories satisfying $r>R_d$ but do not escape at the epoch when $r>R_d$ (e.g., the states of trajectory satisfy $\dot r<0$ after $r>R_d$). Therefore, our works continue to adopt this criterion to identify escape trajectories in the Earth-Moon PCR3BP. The generalized energy $E$ can be expressed as follows \cite{qi2015mechanical}:
\begin{equation}
E = \frac{1}{2}\left( \left(u-y\right)^2+\left(v+x\right)^2 \right) - \frac{{1 - \mu }}{{{r_1}}} - \frac{\mu }{{{r_2}}} \label{eq4}
\end{equation}
We make some refinement to this criterion, and the updated criterion can be expressed as follows:
\begin{enumerate}
\item The value of $r$ exceeds a preset threshold, i.e., $r > R_d$.
\item The value of $\dot r$ satisfies $  \text{d}r/\text{d}t > 0$.
\item The value of $E$ satisfies $ E > 0$.
\item The trajectories do bot impact the Earth and the Moon, i.e., $r_1\geq R_\text{Earth}$ and $r_2\geq R_\text{Moon}$, where $R_\text{Earth}=6378.145 \text{ km}$ and $R_\text{Moon}=1737.100\text{ km}$ denote the radius of the Earth and the Moon.
\end{enumerate}
When the trajectories satisfy the above four conditions simultaneously, the trajectories are identified as escape trajectories in the Earth-Moon PCR3BP. Following Refs. \cite{zotos2015classifying,FU2025,fu2025energy}, we still set $R_d$ to 10 LU. 

When considering escape trajectories in the Sun-Earth/Moon PBCR4BP, the definition of escape trajectories should be further refined because trajectories satisfying $r \to \infty$ when $t \to \infty$ are not entirely equivalent to those escaping from the Earth-Moon system (roughly, this definition in the Sun-Earth/Moon PBCR4BP may be suitable for trajectories escaping from the solar system). We borrow the definition of trajectories escaping from the secondary body in the PCR3BP proposed by Ano{\`e} et al. \cite{anoe2024ballistic} and propose the definition of escape trajectories in the Sun-Earth/Moon PBCR4BP. When $r$ is sufficiently large, the Sun-Earth/Moon PBCR4BP can be approximated as the Sun-(Earth-Moon barycenter) PCR3BP \cite{qi2017research}. Escape trajectories from the Earth-Moon system can be approximated as trajectories escaping from the secondary body in the Sun-(Earth-Moon barycenter) PCR3BP. Therefore, the definition (criterion) can be expressed as follows: when trajectories satisfy $r>R_d$ combined with $E_2>0$ for a finite integration time $t$, trajectories can be identified as escape trajectories in the Sun-Earth/Moon PBCR4BP, where $E_2$ denotes the two-energy with respect to the Earth-Moon barycenter (the definition of the two-energy with respect to the Earth-Moon barycenter can be found in Ref. \cite{FU2025}). Because there exists an approximation relationship between $E$ and $E_2$ when $r$ is sufficiently large \cite{FU2025,fu2025energy}, and we also expect to perform a fair comparison between escape trajectories in the Earth-Moon PCR3BP and the Sun-Earth/Moon PBCR4BP to analyze the effects of the solar gravity perturbation, we adopt the above criterion to identify escape trajectories in the Sun-Earth/Moon PBCR4BP. Subsequently, escape families existing in these two models and their identification method are presented in Section \ref{sec3}. 

\section{Escape-Family Identification Using Data Mining}\label{sec3}
This section proposes an identification method of escape families, including constructing the solution space of escape trajectories, pre-filtering escape trajectories, and identifying escape families using data mining. 

\subsection{Constructing Solution Space of Escape Trajectories}\label{subsec3.1}
\begin{figure}[H]
\centering
\includegraphics[width=0.3\textwidth]{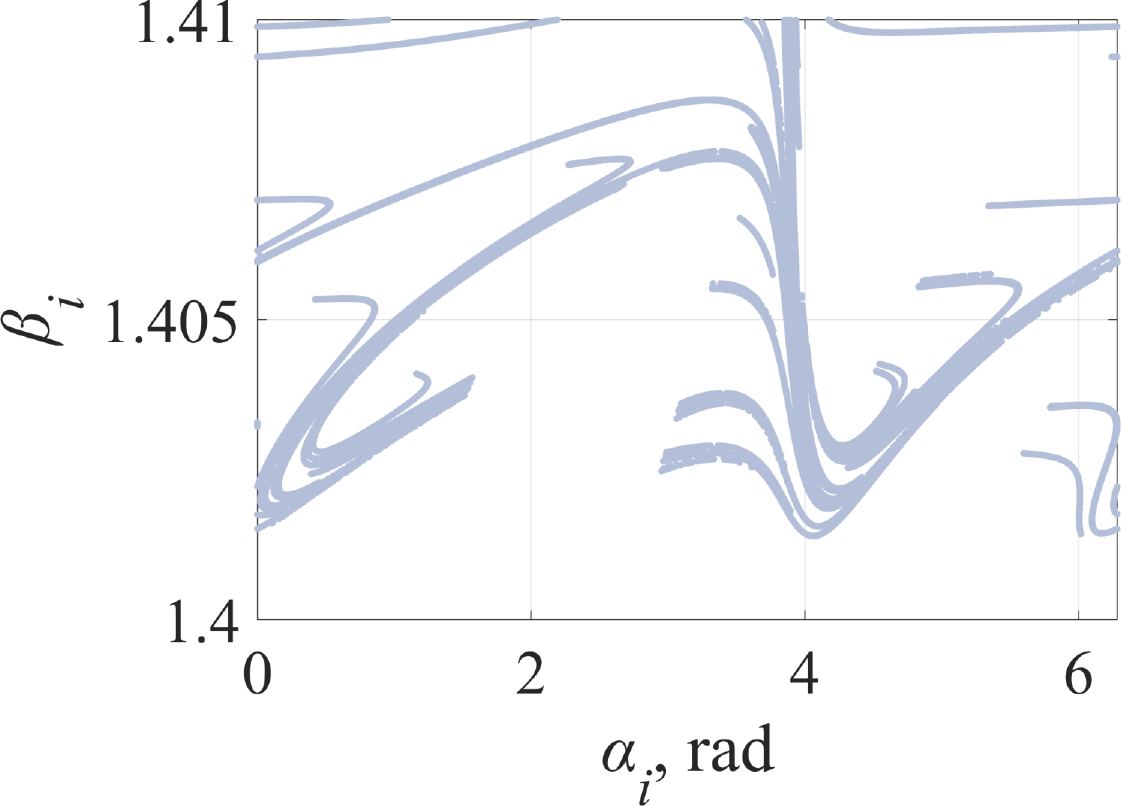}
\caption{The $\left(\alpha_i,\text{ }\beta_i\right)$ map of identified escape trajectories in the Earth-Moon PCR3BP.}
\label{fig_three_map}
\end{figure}

\begin{figure}[h]
\centering
\includegraphics[width=0.6\textwidth]{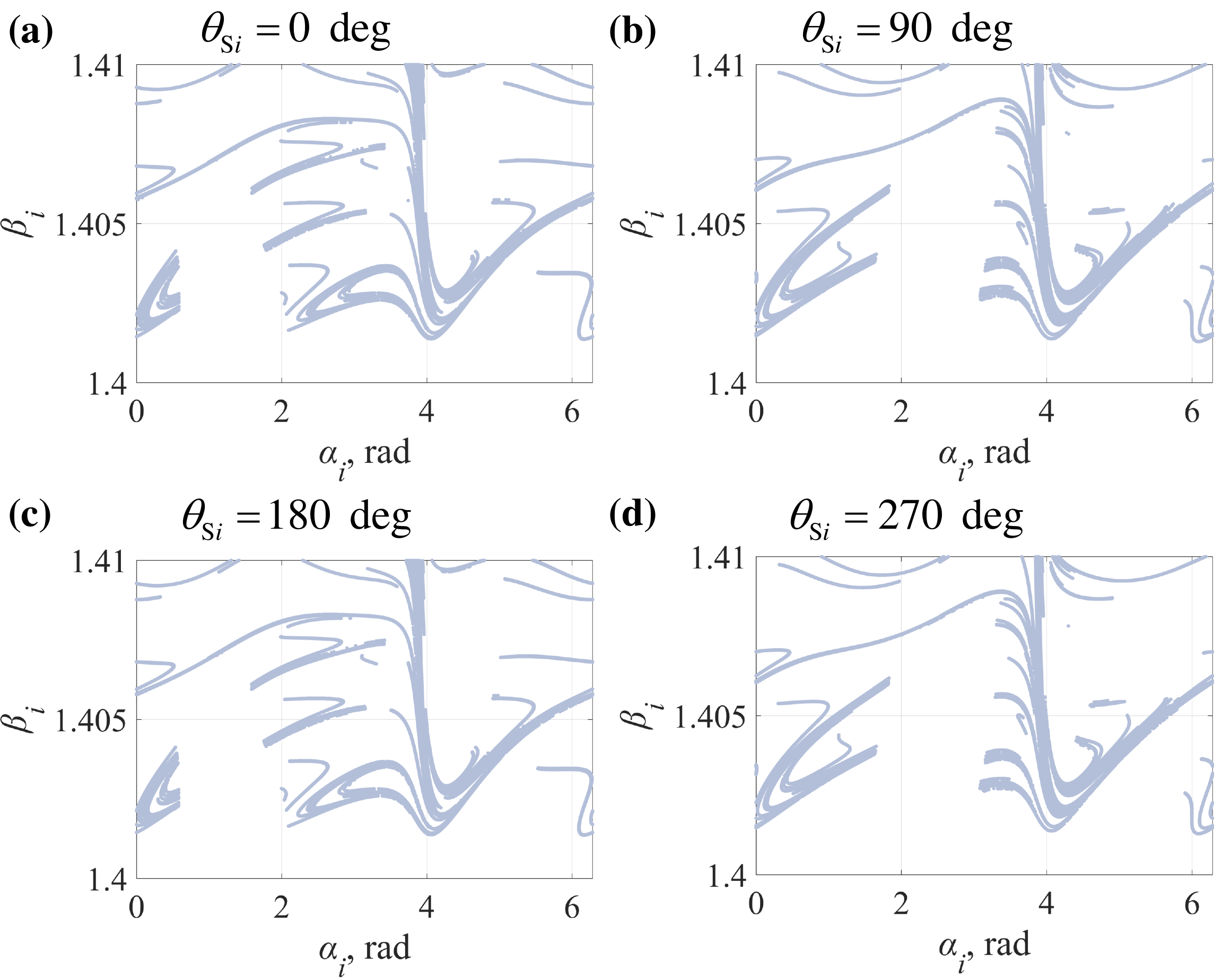}
\caption{The $\left(\alpha_i,\text{ }\beta_i\right)$ map of identified escape trajectories in the Sun-Earth/Moon PBCR4BP. (a) $\theta_{\text{S}i}=0\text{ }\deg$; (b) $\theta_{\text{S}i}=90\text{ }\deg$; (c) $\theta_{\text{S}i}=180\text{ }\deg$; (d) $\theta_{\text{S}i}=270\text{ }\deg$.}
\label{fig_four_map}
\end{figure}

We adopt the grid search method to generate the solution space of escape trajectories in the Earth-Moon PCR3BP and the Sun-Earth/Moon PBCR4BP. We focus on trajectories escaping from a 167 km circular Earth orbit. For this type of escape trajectories, the departure conditions that trajectories should satisfy can be expressed as \cite{fu2025four,fu2025energy}:
\begin{equation}
{{{\left( {{x_i} + \mu } \right)}^2} + {y_i}^2 - {{\left( {{R_{\text{Earth}}} + {h_i}} \right)}^2}}=0\text{ }\text{ }\text{ }{\left( {{x_i} + \mu } \right)\left( {{u_i} - {y_i}} \right) + {y_i}\left( {{v_i} + {x_i} + \mu } \right)}=0 \label{eq_psi}
\end{equation}
where the subscript ‘\textit{i}’ denotes quantities corresponding to the initial epoch (i.e., the epoch when the spacecraft depart from the circular Earth orbit). The following construction parameters are adopted \cite{topputo2013optimal}:
\begin{equation}
\bm{y}=\left[\alpha_i,\text{ }\beta_i\right]^\text{T} \label{eq5}
\end{equation}
where $\alpha_i$ denotes the departure phase angle at the circular Earth orbit, and $\beta_i$ denotes the initial-to-circular velocity ratio. With these parameters, the initial states of trajectories $\bm{X}_i$ can be expressed as follows \cite{topputo2013optimal}:
\begin{equation}
  {x_i} = {r_i}\cos {\alpha _i} - \mu  \text{ }\text{ }\text{ }
  {y_i} = {r_i}\sin {\alpha _i} \text{ }\text{ }\text{ }
  {u_i} =  - \left( {{\beta _i}\sqrt {\frac{{1 - \mu }}{{{r_i}}}}  - {r_i}} \right)\sin {\alpha _i} \text{ }\text{ }\text{ }
  {v_i} = \left( {{\beta _i}\sqrt {\frac{{1 - \mu }}{{{r_i}}}}  - {r_i}} \right)\cos {\alpha _i} 
 \label{eq6}
\end{equation}
where $r_i=h_i+R_\text{Earth}$ ($h_i=167 \text{ km}$). The maximum integration time is set to 90 days for the practical mission consideration. When the trajectory satisfy the escape criterion mentioned in Section \ref{subsec2.2}, the numerical integration terminates and the time of flight (TOF) is recorded. The TOF is calculated by $\text{TOF}=t_f-t_i$, where $t_f$ denotes the termination epoch and $t_i$ denotes the initial epoch ($t_i=0$ for the Earth-Moon PCR3BP and $t_i=\theta_{\text{S}i}/\omega_\text{S}$ for the Sun-Earth/Moon PBCR4BP). The Earth injection impulse ($\Delta v_i$) of the escape trajectories can be calculated by:
\begin{equation}
\Delta {v_i} = \sqrt {{{\left( {{u_i} - {y_i}} \right)}^2} + {{\left( {{v_i} + {x_i} + \mu } \right)}^2}}  - \sqrt {\frac{{1 - \mu }}{r_i }} =\left(\beta_i-1\right)\sqrt {\frac{{1 - \mu }}{r_i }} \label{eq_vi}
\end{equation}
To exhaustively search escape trajectories satisfying the criterion mentioned in Section \ref{subsec2.2} within 90 days, the grid search method is adopted. Referring to Ref. \cite{topputo2013optimal}, we set $\alpha_i$ as $\alpha_i \in \left[0,\text{ }2\pi\right)$ with a step-size of $\pi/7200$, and set $\beta_i$ as $\beta_i \in \left[1.4,\text{ }1.41\right]$ with a step-size of 0.000002. For the Sun-Earth/Moon PBCR4BP, we set $\theta_{\text{S}i}=0,\text{ }90,\text{ }180,\text{ }270\text{ }\deg$. To perform a fair comparison between escape trajectories in the Earth-Moon PCR3BP and the Sun-Earth/Moon PBCR4BP, we particularly investigate escape trajectories generated by the same initial states for each case. The corresponding $\left(\alpha_i,\text{ }\beta_i\right)$ map of escape trajectories in the Earth-Moon PCR3BP is presented in Fig. \ref{fig_three_map}, and corresponding $\left(\alpha_i,\text{ }\beta_i\right)$ map in the Sun-Earth/Moon PBCR4BP for four cases ($\theta_{\text{S}i}=0,\text{ }90,\text{ }180,\text{ }270\text{ }\deg$) is presented in Fig. \ref{fig_four_map}.

From Figs. \ref{fig_three_map}-\ref{fig_four_map}, it is shown that $\left(\alpha_i,\text{ }\beta_i\right)$ exhibits a clustered distribution pattern on the map. It can be also observed that the $\left(\alpha_i,\text{ }\beta_i\right)$ distributions for the Earth-Moon PCR3BP and the Sun-Earth/Moon PBCR4BP can be different. These phenomenon motivate us to use data mining techniques to analyze the escape trajectories and further explore the effects of solar gravity perturbation. Meanwhile, in Fig. \ref{fig_four_map}, we find that the $\left(\alpha_i,\text{ }\beta_i\right)$ distributions are similar to cases where the difference between $\theta_{\text{S}i}$ is 180 deg, which implies that the dynamical behavior of trajectories is similar with the values of $\theta_{\text{S}i}$ differing 180 deg (the similar phenomenon can be also found in Refs. \cite{ren2014low,FU20254993,fu2025four}). Due to this phenomenon, we only focus on the escape-family identification for $\theta_{\text{S}i}=0\text{ }\deg$ and $\theta_{\text{S}i}=90\text{ }\deg$. Before applying data mining techniques, pre-filter of identified escape trajectories based on dynamical analysis is performed to ensure the performance of the identification method using the data mining techniques.
\subsection{Pre-Filtering Escape Trajectories}\label{subsec3.2}
\begin{figure}[h]
\centering
\includegraphics[width=0.2\textwidth]{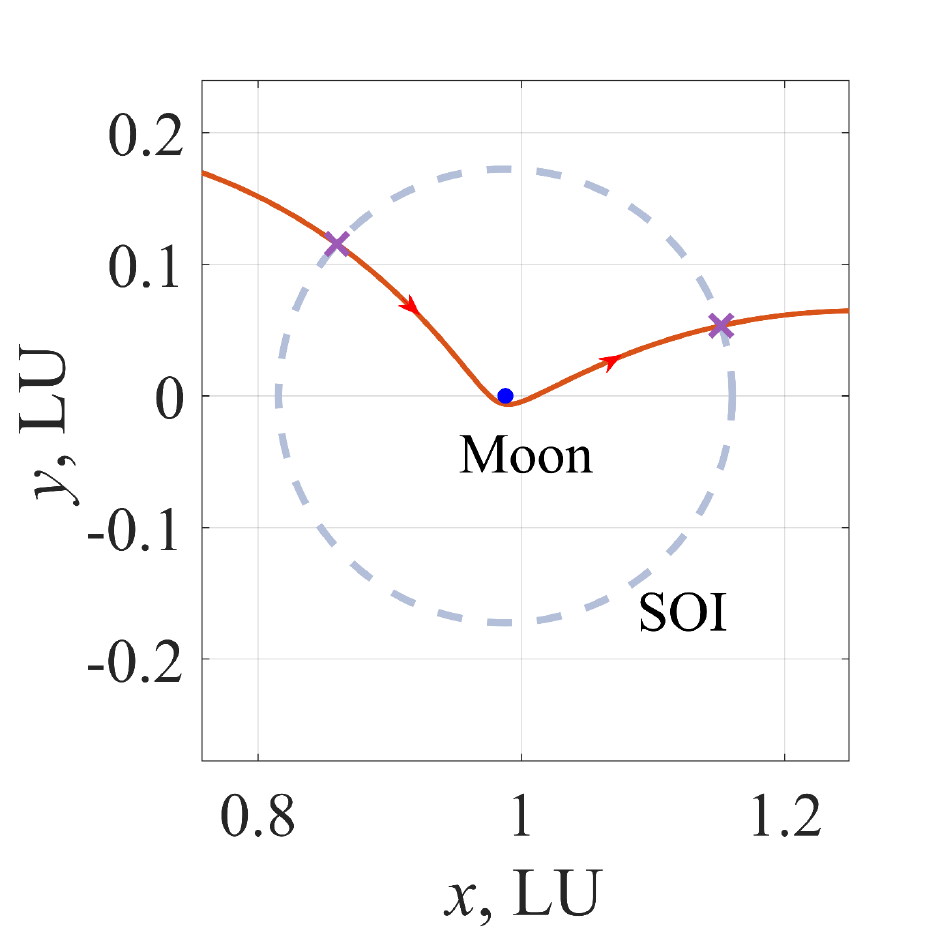}
\caption{The schematic of LGA.}
\label{fig_LGA}
\end{figure}

\begin{figure}[h]
\centering
\includegraphics[width=0.9\textwidth]{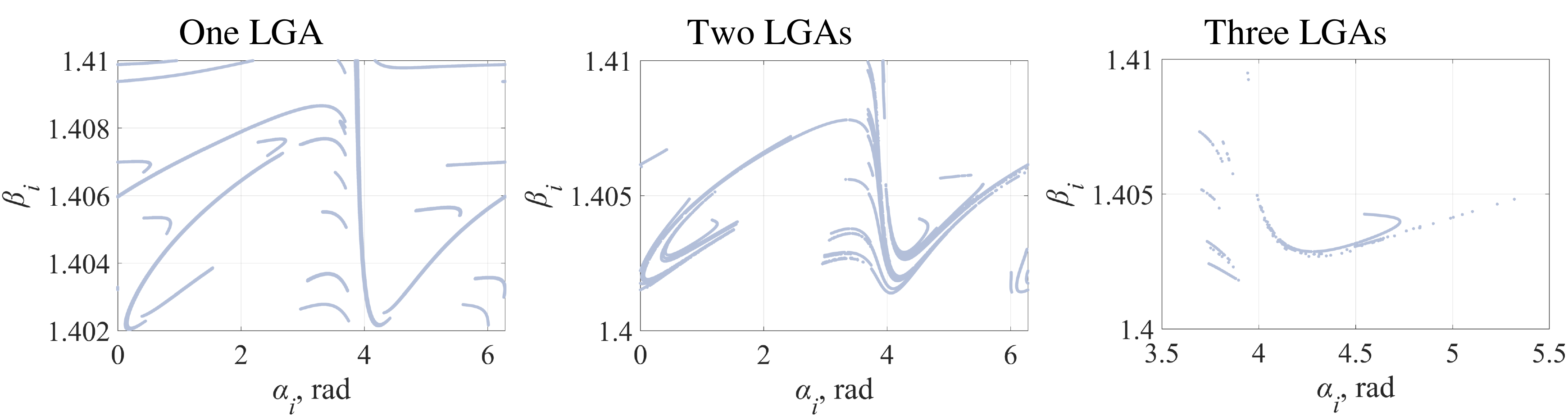}
\caption{The $\left(\alpha_i,\text{ }\beta_i\right)$ map of identified escape trajectories with LGAs in the Earth-Moon PCR3BP.}
\label{fig_3_LGA}
\end{figure}

\begin{figure}[h]
\centering
\includegraphics[width=0.9\textwidth]{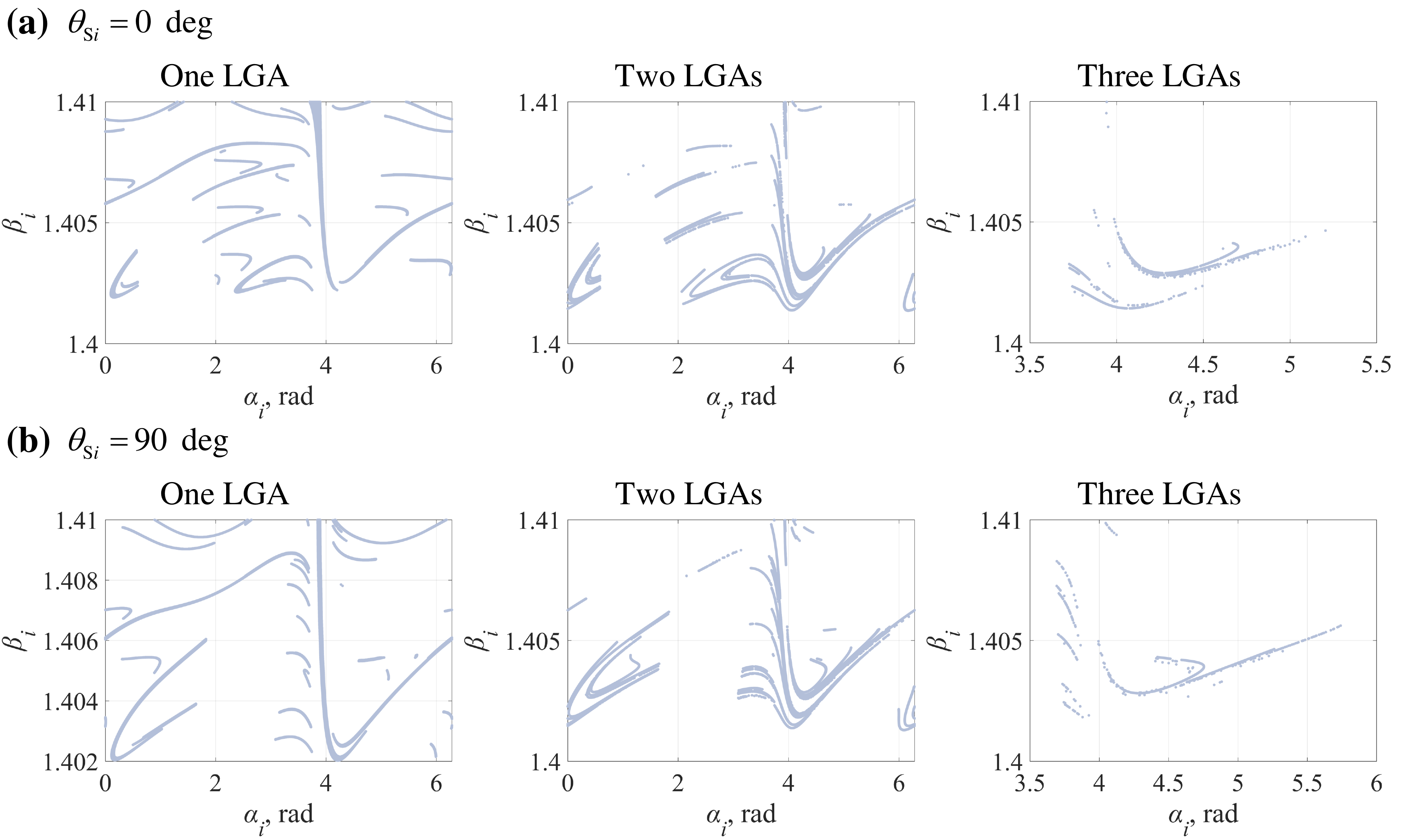}
\caption{The $\left(\alpha_i,\text{ }\beta_i\right)$ map of identified escape trajectories with LGAs in the Sun-Earth/Moon PBCR4BP. (a) $\theta_{\text{S}i}=0\text{ }\deg$; (b) $\theta_{\text{S}i}=90\text{ }\deg$.}
\label{fig_4_LGA}
\end{figure}

In this subsection, we perform a pre-filter of the obtained escape trajectories based on their dynamical characteristics. According to Refs. \cite{qi2015mechanical,FU2025,fu2025energy}, LGA plays an important role in the variation in $E$. Therefore, we use the number of LGA to pre-filter the escape trajectories. We define escape trajectories with LGAs as escape trajectories intersecting the disk of sphere of influence (SOI) of the Moon (two intersections for one LGA, four intersections for two LGAs, and so on). The radius of SOI of the Moon is set to 66243 km \cite{pan2022research}. A schematic of trajectory with one LGA can be found in Fig. \ref{fig_LGA}, and x-shaped markers denote the intersection of the disk of SOI of the Moon. With the parameter setting mentioned in Section \ref{subsec3.1}, there is no escape trajectory without LGA both for the Earth-Moon PCR3BP and the Sun-Earth/Moon PBCR4BP. The maximum number of LGA both for the Earth-Moon PCR3BP and the Sun-Earth/Moon PBCR4BP is 3. The corresponding $\left(\alpha_i,\text{ }\beta_i\right)$ maps of escape trajectories with LGAs are presented in Figs. \ref{fig_3_LGA}-\ref{fig_4_LGA}. Moreover, the numbers of trajectories with one LGA, two LGAs, and three LGAs are summarized in Table \ref{tab1}. From these figures and table, it can be observed that for each number of LGA, the number of escape trajectories in the Sun-Earth/Moon PBCR4BP (both two values of $\theta_{\text{S}i}$) is larger than those in the Earth-Moon PCR3BP. Meanwhile, escape trajectories with one LGA accounts for the majority of all escape trajectories, and the initial states of these trajectories exhibit a more clustered distribution than those with two and three LGAs. Also, from a perspective of transfer characteristics, we found that the minimum $\Delta v_i$ of escape trajectories with one LGA is comparable to that of all escape trajectories for these three cases (shown in Table \ref{tab_new}). Therefore, we consider the escape trajectories with one LGA as the typical escape trajectories, and perform the analysis on these trajectories. Subsequently, the identification method is detailed in the next subsection.

\begin{table}[h]
\centering
\renewcommand{\arraystretch}{1.5}
\caption{Numbers of escape trajectories with one LGA, two LGAs, and three LGAs.}\label{tab1}%
\begin{tabular}{@{}llll@{}}
\hline
Model    & Number of LGA & $\theta_{\text{S}i}$, deg & Number of Escape Trajectories \\
\hline
Earth-Moon PCR3BP & 1  & -- & 797197 \\
 & 2  & -- & 68151 \\
  & 3  & -- & 622 \\
Sun-Earth/Moon PBCR4BP & 1  & 0 & 1280411 \\
 & 2  & 0 & 75521 \\
 & 3  & 0 & 1075 \\
 Sun-Earth/Moon PBCR4BP & 1  & 90 & 1058458 \\
 & 2  & 90 & 85963 \\
 & 3  & 90 & 954 \\
\hline
\end{tabular}
\end{table}

\begin{table}[h]
\centering
\renewcommand{\arraystretch}{1.5}
\caption{The minimum $ \Delta v_i$ of escape trajectories with one LGA and all escape trajectories.}\label{tab_new}%
\begin{tabular}{@{}llll@{}}
\hline
Model    & $\theta_{\text{S}i}$, deg & $\min \Delta v_i$ of Trajectories with One LGA, km/s & $\min \Delta v_i$ of All Trajectories, km/s \\
\hline
Earth-Moon PCR3BP & -- & 3.133412 & 3.128439 \\
Sun-Earth/Moon PBCR4BP  & 0 & 3.132461 & 3.128190 \\
 Sun-Earth/Moon PBCR4BP   & 90 & 3.133349 & 3.127644 \\
\hline
\end{tabular}
\end{table}

\subsection{Identifying Escape Families Using Data Mining}\label{subsec3.3}
\begin{figure}[h]
\centering
\includegraphics[width=0.9\textwidth]{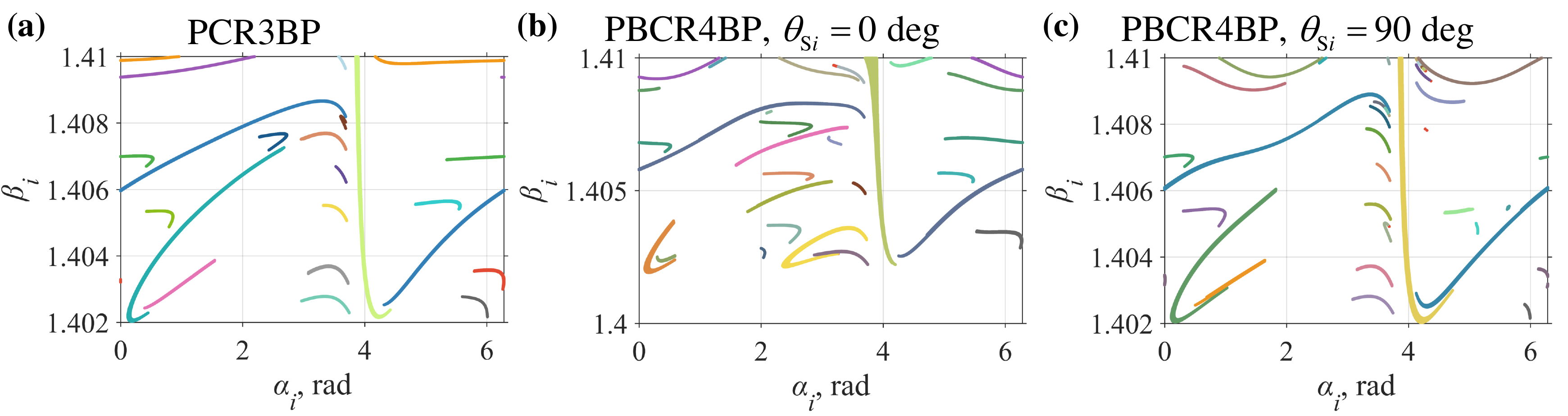}
\caption{The identification results on $\left(\alpha_i,\text{ }\beta_i\right)$ map. (a) Earth-Moon PCR3BP; (b)Sun-Earth/Moon PBCR4BP, $\theta_{\text{S}i}=0\text{ }\deg$; (c)Sun-Earth/Moon PBCR4BP, $\theta_{\text{S}i}=90\text{ }\deg$.}
\label{fig_cluster}
\end{figure}

In this subsection, the escape-family identification method is proposed based on data mining techniques. Based on the above dynamical analysis in Section \ref{subsec3.2}, we focus on the identification of escape families with one LGA. Moreover, based on the $\left(\alpha_i,\text{ }\beta_i\right)$ maps shown in Figs. \ref{fig_3_LGA}-\ref{fig_4_LGA}, the DBSCAN algorithm is utilized because it can identify clusters of arbitrary shapes and does not required to the number of clusters to be specified in advance \cite{raschka2022machine}. The compressed description vector can be selected as follows:
\begin{equation}
{\bm{Y}} = \left[\sin \alpha_i,\text{ }  \cos \alpha_i,\text{ } \hat \beta_i\right]^\text{T}\label{eq_vector}
\end{equation}
where $\hat \beta_i$ denotes the $\beta_i$ normalized into $\left[0,\text{ }1\right]$:
\begin{equation}
\hat \beta_i = \frac{\beta_i-\min \beta_i}{\max\beta_i-\min \beta_i}\label{eq_beta}
\end{equation}
In the compressed description vector, $\beta_i$ is normalized due to the different physical insight from $\alpha_i$, and $\sin \alpha_i$ and $\cos \alpha_i$ are adopted instead of $\alpha_i$ because the values of $\alpha_i$ close to $0$ and $2\pi$ are physically close. Then, the DBSCAN algorithm is performed. The Euclidean distance \cite{bosanac2020data} is adopted for the similarity measure. There exists two hyperparameters of the DBSCAN algorithm, namely, $\varepsilon$ and $minPts$. The hyperparameter $\varepsilon$ denotes the maximum neighborhood distance (i.e., the radius of the $\varepsilon$-neighborhood) of a point, and $minPts$ denotes the minimum number of points within an $\varepsilon$-neighborhood of a point. The setting of these two hyperparameters is determined through repeated trials to effectively identify cluster structures (i.e., escape families). The identification results and the corresponding settings of hyperparameters are presented in Fig. \ref{fig_cluster} and Table \ref{tab2}. For each subfigure in Fig. \ref{fig_cluster}, different colors denote the different escape families, verifying the effectiveness of the proposed identification method. Notably, The same family, characterized by similar $\left(\alpha_i,\text{ }\beta_i\right)$ distributions, may be represented by different colors in subfigures (a)–(c). With the hyperparameter setting presented in Table \ref{tab2}, 19 families are identified for the Earth-Moon PCR3BP, 24 families are identified for the Sun-Earth/Moon PBCR4BP with $\theta_{\text{S}i}=0\text{ }\deg$, and 32 families are identified for the Sun-Earth/Moon PBCR4BP with $\theta_{\text{S}i}=90\text{ }\deg$. There are no noise points for the Earth-Moon PCR3BP. Moreover, when considering the Sun-Earth/Moon PBCR4BP, the numbers of noise points are 15 for $\theta_{\text{S}i}=0\text{ }\deg$ and 11 for $\theta_{\text{S}i}=90\text{ }\deg$, respectively. The detailed analysis on the identified escape families is performed in Section \ref{sec4}.
\begin{table}[h]
\centering
\renewcommand{\arraystretch}{1.5}
\caption{The setting of hyperparameters of the DBSCAN algorithm.}\label{tab2}%
\begin{tabular}{@{}llll@{}}
\hline
Model     & $\theta_{\text{S}i}$, deg & $minPts$ & $\varepsilon$ \\
\hline
Earth-Moon PCR3BP   & -- & 25 & 0.018 \\
Sun-Earth/Moon PBCR4BP   & 0 & 4 & 0.008 \\
Sun-Earth/Moon PBCR4BP   & 90 & 4 & 0.008 \\
\hline
\end{tabular}
\end{table}

\section{Results and Discussion} \label{sec4}
This section presents the detailed analysis on the identification results of the proposed method in Section \ref{sec3}. Escape families in the Earth-Moon PCR3BP and the Sun-Earth/Moon PBCR4BP are presented and analyzed. Based on the presented results, the effects of the solar gravity perturbation are further analyzed.
\subsection{Escape Families in The Earth-Moon PCR3BP}\label{subsec4.1}
\begin{figure}[H]
\centering
\includegraphics[width=0.75\textwidth]{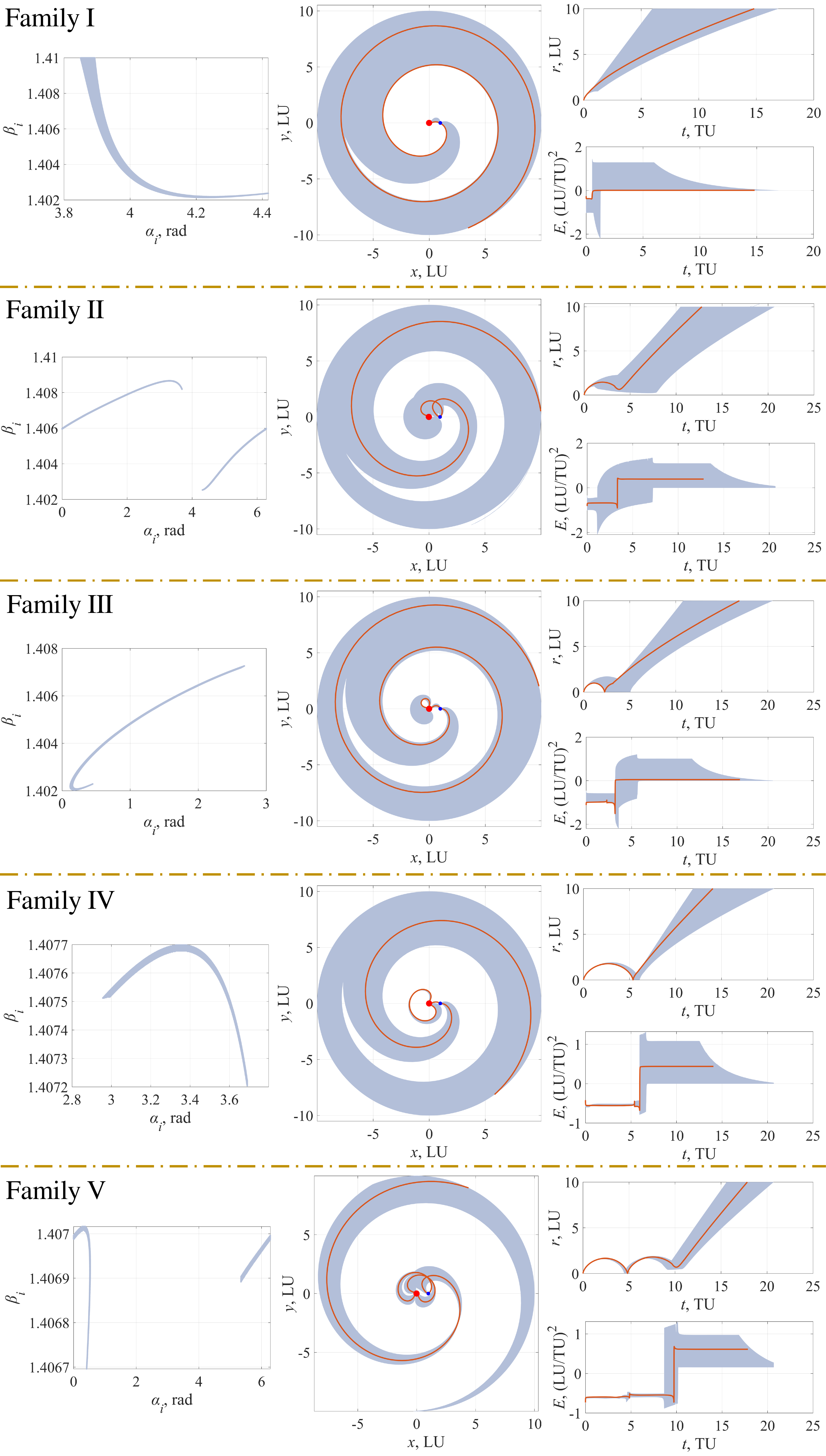}
\caption{The identified escape families in the Earth-Moon PCR3BP (Families I-V).}
\label{fig_3_family_1}
\end{figure}

\begin{figure}[H]
\centering
\includegraphics[width=0.75\textwidth]{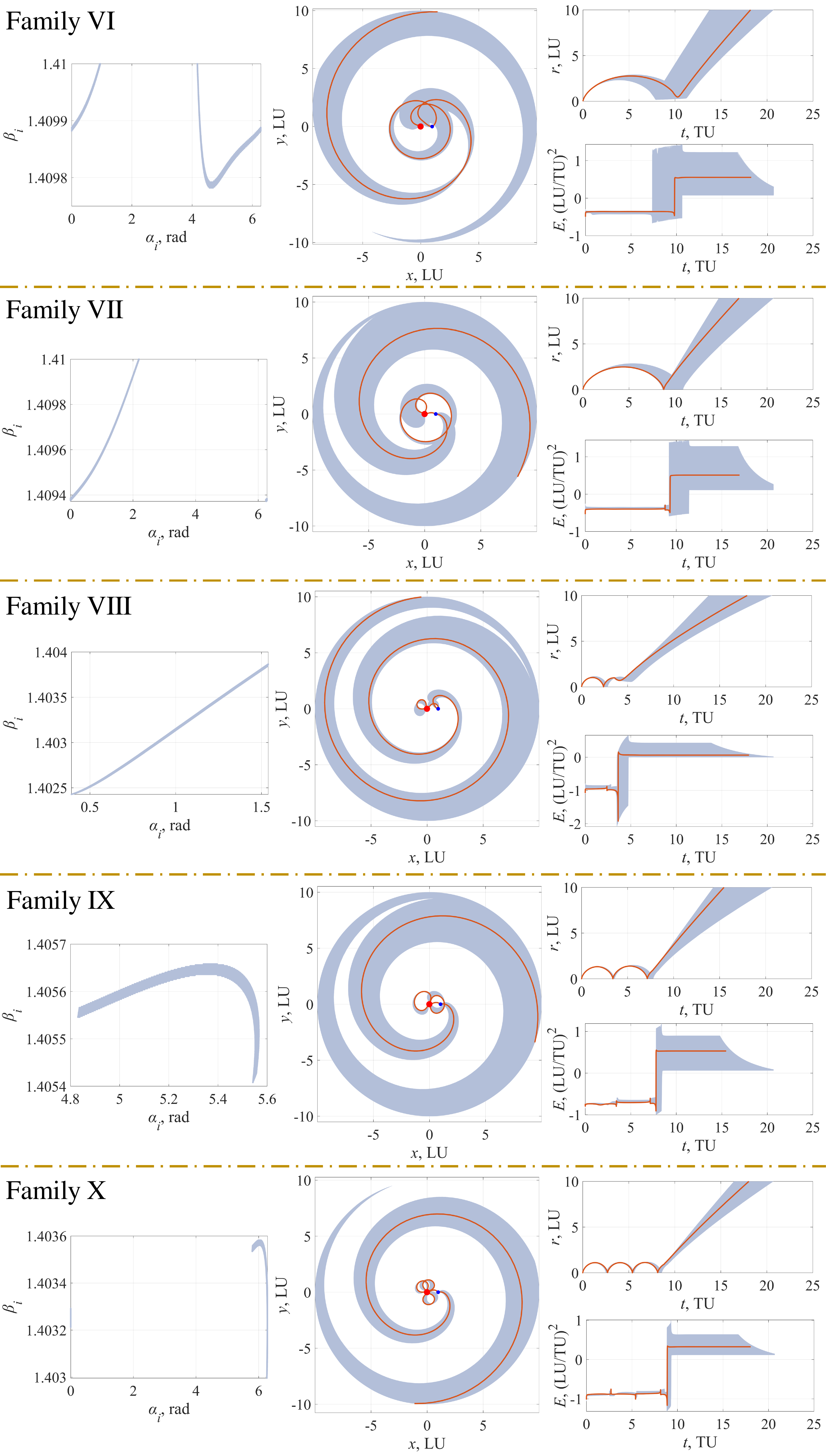}
\caption{The identified escape families in the Earth-Moon PCR3BP (Families VI--X).}
\label{fig_3_family_2}
\end{figure}

\begin{figure}[H]
\centering
\includegraphics[width=0.75\textwidth]{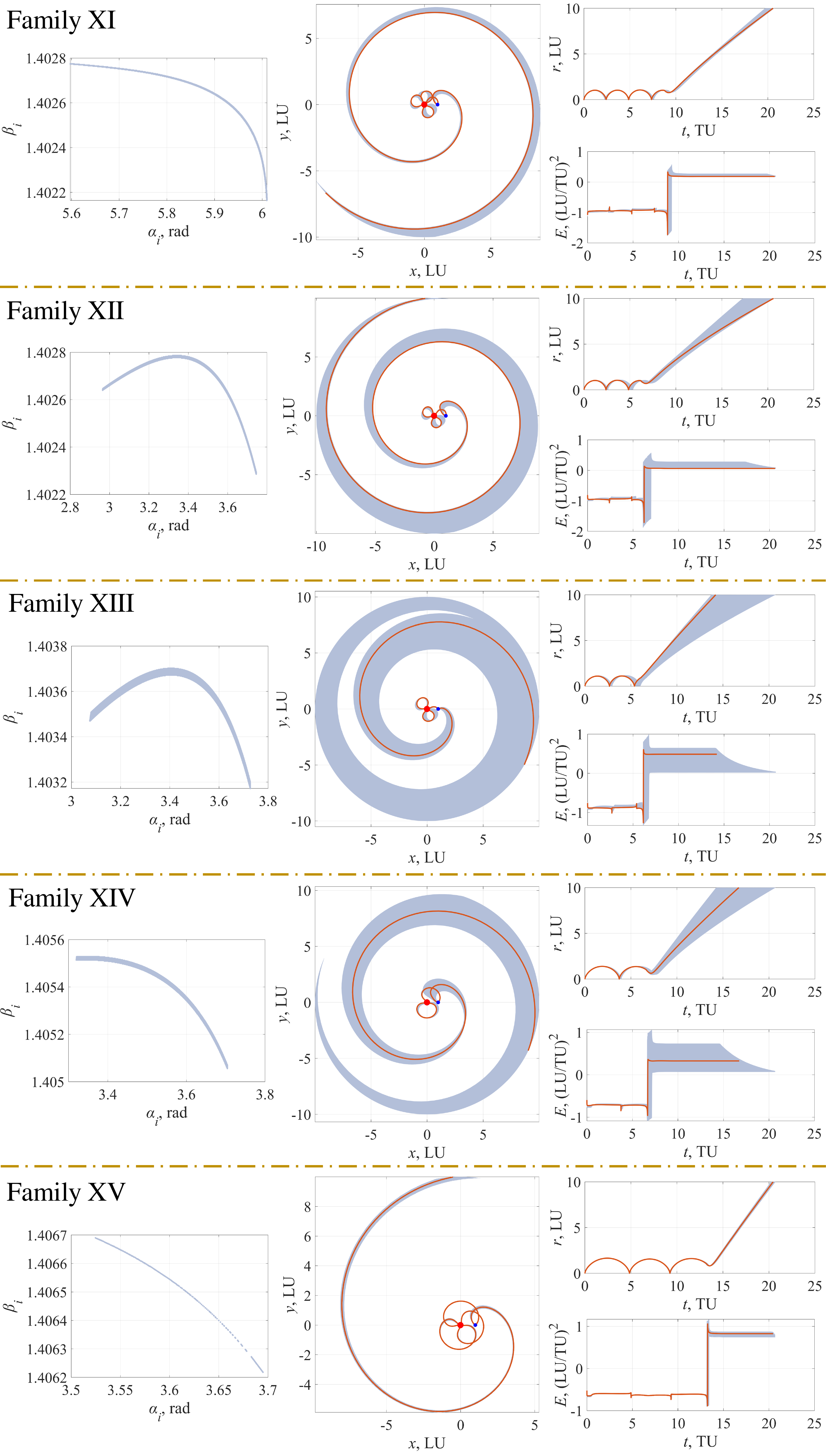}
\caption{The identified escape families in the Earth-Moon PCR3BP (Families XI--XV).}
\label{fig_3_family_3}
\end{figure}

\begin{figure}[H]
\centering
\includegraphics[width=0.75\textwidth]{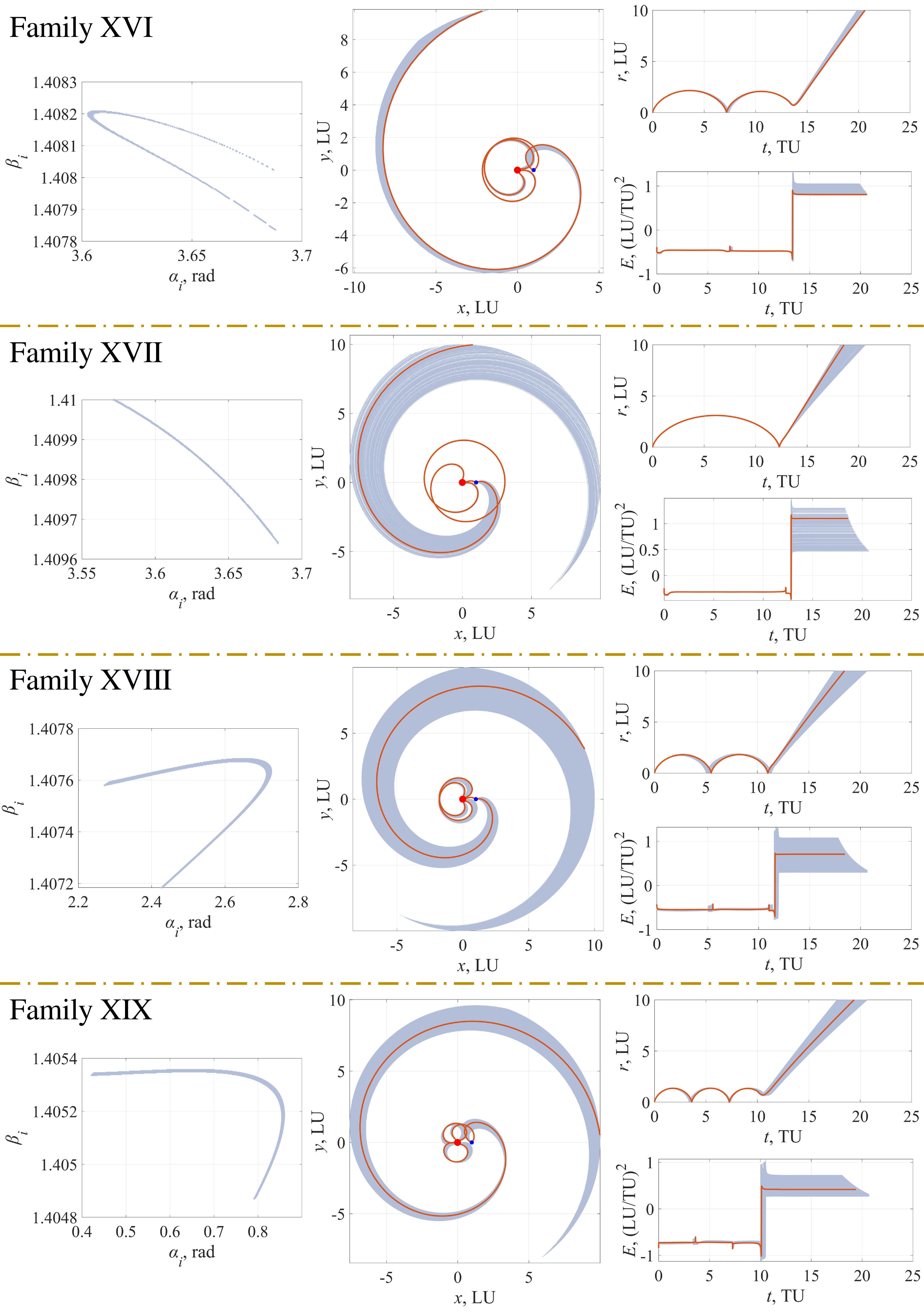}
\caption{The identified escape families in the Earth-Moon PCR3BP (Families XVI--XIX).}
\label{fig_3_family_4}
\end{figure}

\begin{figure}[H]
\centering
\includegraphics[width=0.6\textwidth]{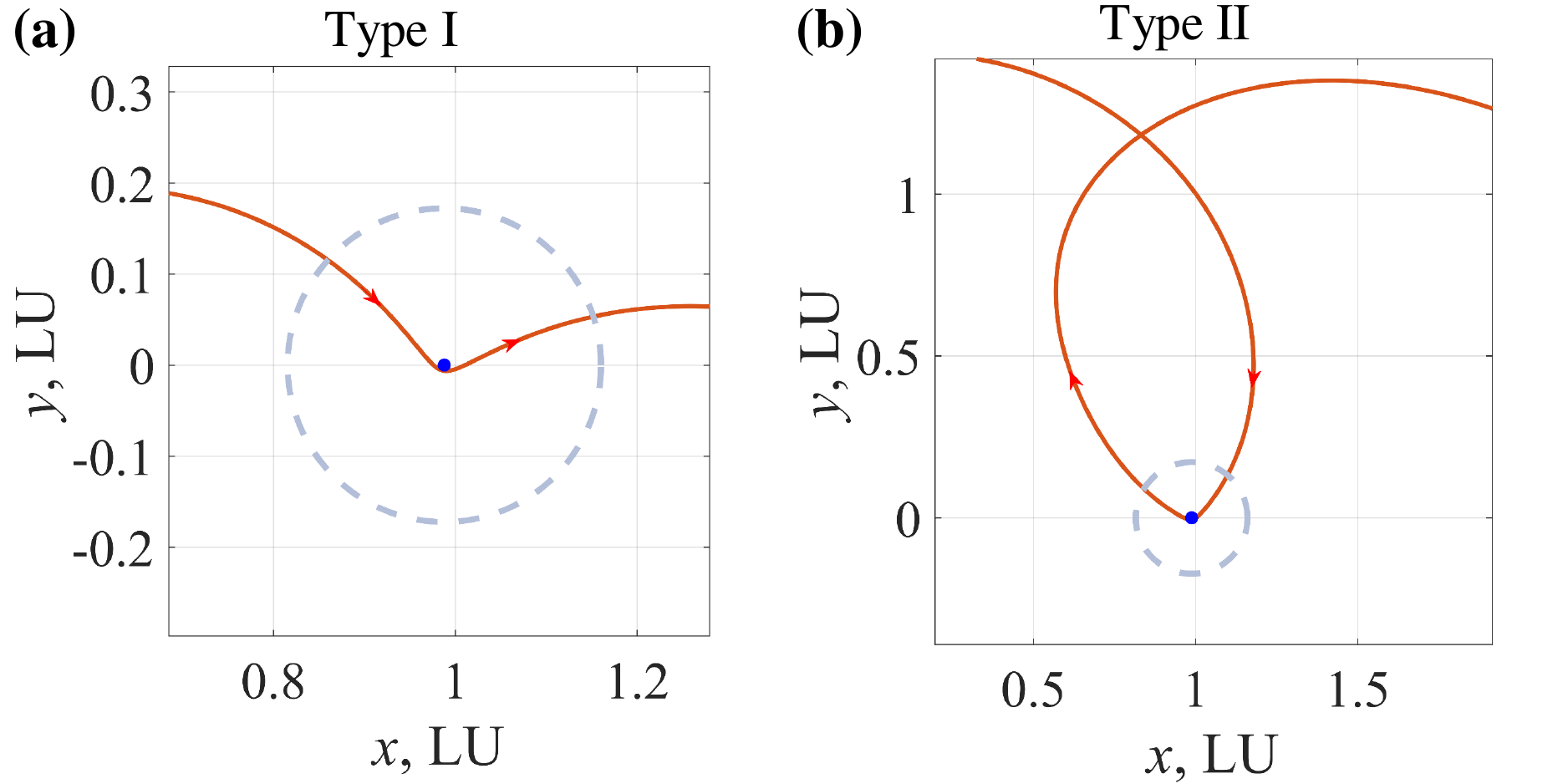}
\caption{Two types of LGA. (a) Prograde LGA (Type I); (b) Retrograde LGA (Type II).}
\label{fig_LGA_type}
\end{figure}

For the Earth-Moon PCR3BP, 19 escape families are identified using the proposed method, shown in Figs. \ref{fig_3_family_1}-\ref{fig_3_family_4}. These 19 families are labeled as Families I-XIX. The corresponding number of escape trajectories for each family is summarized in Table \ref{tab3}. It is found that the number of escape trajectories belonging to Families I-III (combined with Family IV) accounts for the majority of all escape trajectories with one LGA. The first column of Figs. \ref{fig_3_family_1}-\ref{fig_3_family_4} presents the $\left(\alpha_i,\text{ }\beta_i\right)$ map of the corresponding family, the second column presents the escape trajectories, and the third column presents the corresponding time history of $r$ and $E$. Red curves denote the typical trajectory of each family and the corresponding time history of $r$ and $E$. Although the escape families are identified based on the $\left(\alpha_i,\text{ }\beta_i\right)$ distributions, the patterns of trajectories in the same escape family are similar. Notably, Family IV can be considered as a continuation of Family III because of the similar trajectory patterns. The discontinuous distribution in the $\left(\alpha_i,\text{ }\beta_i\right)$ map between Families III and IV is possibly because of the constrained TOF ($\max \text{TOF}=90\text{ days}$) and the constraint of one LGA. Similar phenomenon can be found for Families VII and XVII. The reason for this discontinuous distribution may be the constrained $\beta_i$. From these figures, it can be observed that the LGA can be categorized into two type: prograde LGA (Type I) and retrograde LGA (Type II), shown in Fig. \ref{fig_LGA_type}, which is similar to the finding from Ref. \cite{qi2016study}. Based on the second column of Figs. \ref{fig_3_family_1}-\ref{fig_3_family_4}, escape families with the prograde LGA are Families I, III, IV, VII, IX, X, XIII, XVII, and XVIII; while escape families with the retrograde LGA are Families II, V, VI, VIII, XI, XII, XIV, XV, XVI, and XIX. From the time history of $E$ shown in the third column of these figures, both prograde and retrograde LGAs can cause a significant variation in $E$. For the Earth-Moon PCR3BP, the value of $E$ remain constant approximately because $\text{d}E/\text{d}t \approx0$ in the region far from the Earth and the Moon \cite{qi2015mechanical,wang2025mechanism}.

\begin{figure}[H]
\centering
\includegraphics[width=0.75\textwidth]{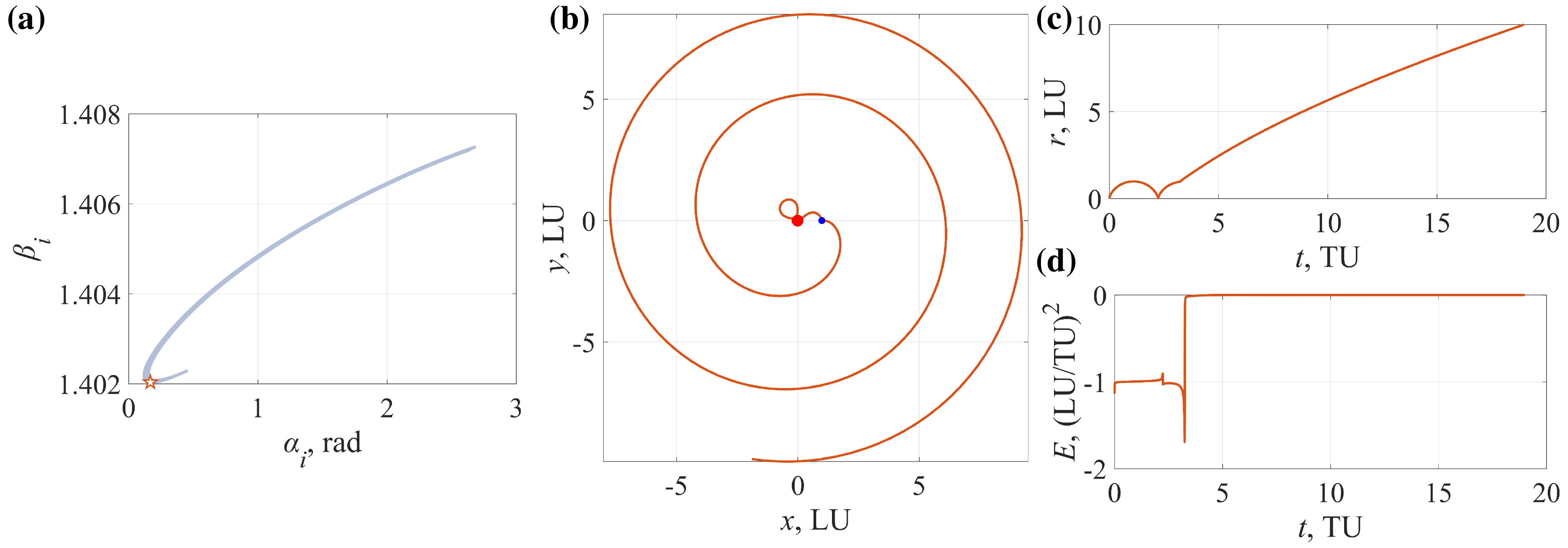}
\caption{Escape trajectory with the minimum $\Delta v_i$ and the corresponding information in the Earth-Moon PCR3BP. (a) Initial states; (b) Trajectory; (c) Time history of $r$; (d) Time history of $E$.}
\label{fig_3_min_V}
\end{figure}

Once the escape families are identified, the transfer characteristics are focused on for potential engineering applications. The initial states of escape trajectories are presented in the first column of Figs. \ref{fig_3_family_1}-\ref{fig_3_family_4}, providing further insights into the selection of initial states in the construction of escape trajectories. The corresponding ranges of $\Delta v_i$ and TOF of these 19 families are summarized in Table \ref{tab3}. From Table \ref{tab3}, it is shown that the escape trajectory with the minimum TOF belongs to Family I, whose TOF is 26 days, while escape trajectory with the minimum $\Delta v_i$ belongs to Family III, whose $\Delta v_i$ is 3.133412 km/s. We focus on the escape trajectories with low $\Delta v_i$ in particular. The escape trajectory with the minimum $\Delta v_i$ and the corresponding information are presented in Fig. \ref{fig_3_min_V}. The red star in Fig. \ref{fig_3_LGA} (a) denote the $\left(\alpha_i,\text{ }\beta_i\right)$ of the trajectory. Subsequently, to reveal the effects of the solar gravity perturbation on the escape trajectories, we present the identification of escape families in the Sun-Earth/Moon PBCR4BP, and focus on the differences between results in the Earth-Moon PBCR3BP and the Sun-Earth/Moon PBCR4BP.


\begin{table}[h]
\centering
\renewcommand{\arraystretch}{1.5}
\caption{Characteristics of identified escape families in the Earth-Moon PCR3BP.}\label{tab3}%
\begin{tabular}{@{}llllll@{}}
\hline
Family & Number & $\min \Delta v_i$, km/s & $\max \Delta v_i$, km/s & $\min\text{TOF}$, Day  & $\max\text{TOF}$, Day \\
\hline
I    & 262374 & 3.134254 & 3.195498 & 26 & 74 \\
II   & 174527 & 3.137262 & 3.185132 & 46 & 90 \\
III  & 153872 & 3.133412 & 3.174158 & 47 & 89 \\
IV   & 18026  & 3.173706 & 3.177556 & 52 & 90 \\
V    & 20203  & 3.169747 & 3.172241 & 68 & 90 \\
VI   & 29979  & 3.193799 & 3.195498 & 60 & 90 \\
VII  & 19561  & 3.190603 & 3.195498 & 64 & 90 \\
VIII & 23198  & 3.136483 & 3.147690 & 60 & 90 \\
IX   & 21758  & 3.159708 & 3.161657 & 62 & 90 \\
X    & 17694  & 3.140910 & 3.145477 & 71 & 90 \\
XI   & 2156   & 3.134425 & 3.139179 & 86 & 90 \\
XII  & 8325   & 3.135376 & 3.139273 & 75 & 90 \\
XIII & 23160  & 3.142281 & 3.146412 & 60 & 90 \\
XIV  & 7041   & 3.156965 & 3.160644 & 62 & 90 \\
XV   & 211    & 3.166021 & 3.169700 & 88 & 90 \\
XVI  & 474    & 3.178632 & 3.181531 & 86 & 90 \\
XVII & 392    & 3.192692 & 3.195498 & 79 & 90 \\
XVIII& 7874   & 3.173550 & 3.177432 & 76 & 90 \\
XIX  & 6372   & 3.155515 & 3.159303 & 77 & 90 \\
\hline
\end{tabular}
\end{table}
\subsection{Escape Families in The Sun-Earth/Moon PBCR4BP ($\theta_{\text{S}i}=0\text{ }\deg$)}\label{subsec4.2}
\begin{figure}[H]
\centering
\includegraphics[width=0.75\textwidth]{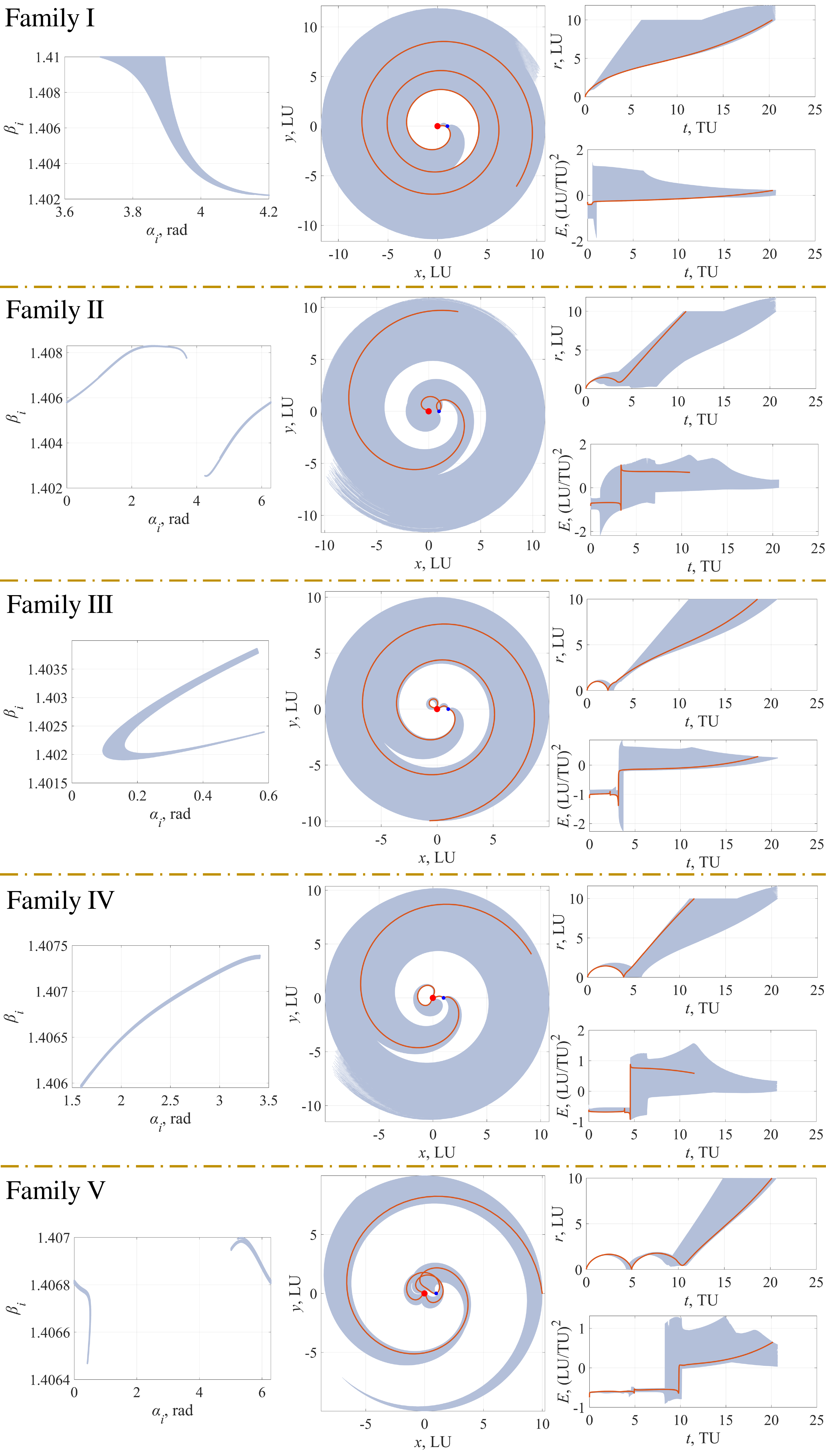}
\caption{The identified escape families in Sun-Earth/Moon PBCR4BP with $\theta_{\text{S}i}=0\text{ }\deg$ (Families I--V).}
\label{fig_4_family_0_deg_1}
\end{figure}

\begin{figure}[H]
\centering
\includegraphics[width=0.75\textwidth]{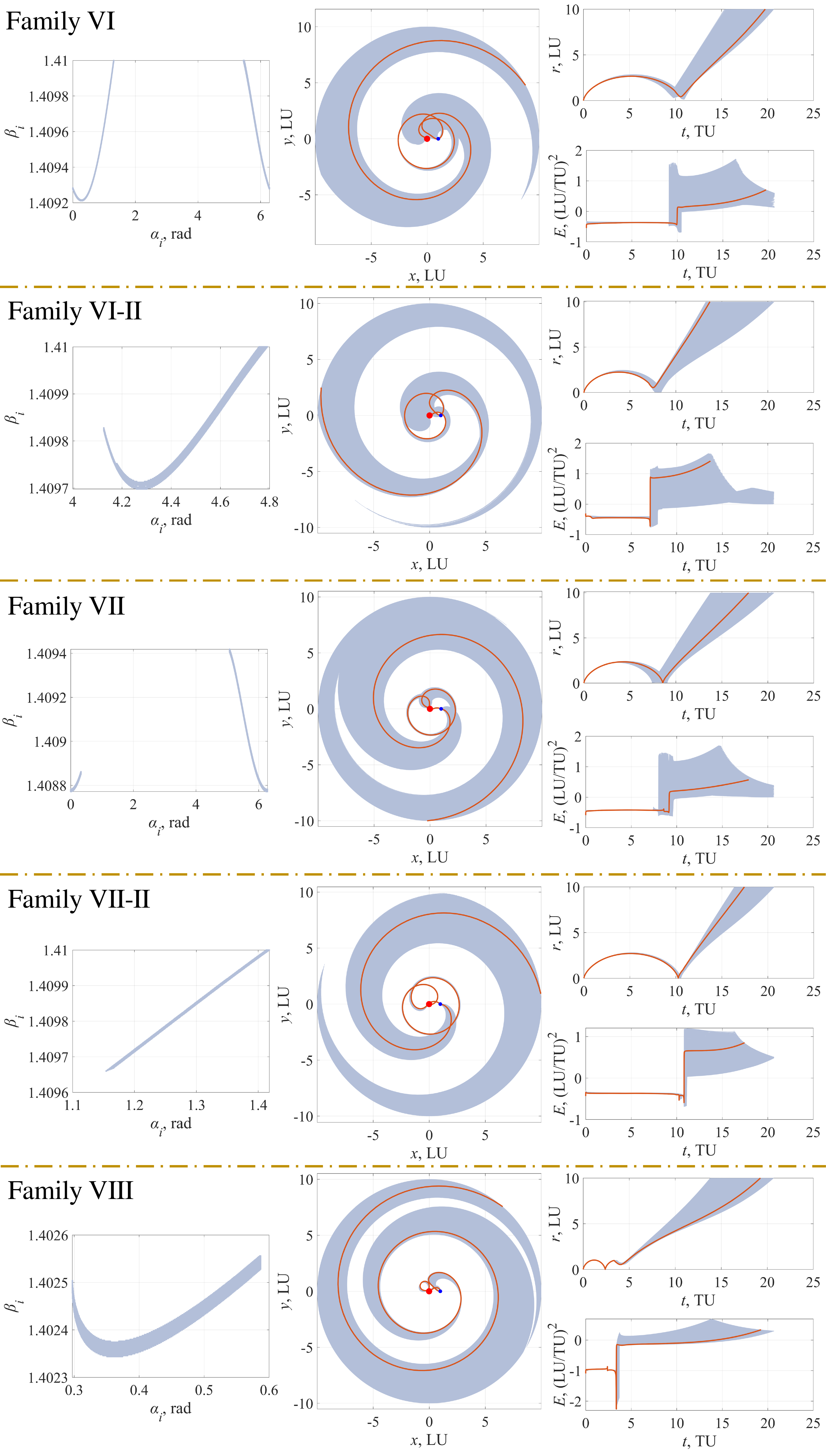}
\caption{The identified escape families in Sun-Earth/Moon PBCR4BP with $\theta_{\text{S}i}=0\text{ }\deg$ (Families VI--VIII).}
\label{fig_4_family_0_deg_2}
\end{figure}

\begin{figure}[H]
\centering
\includegraphics[width=0.75\textwidth]{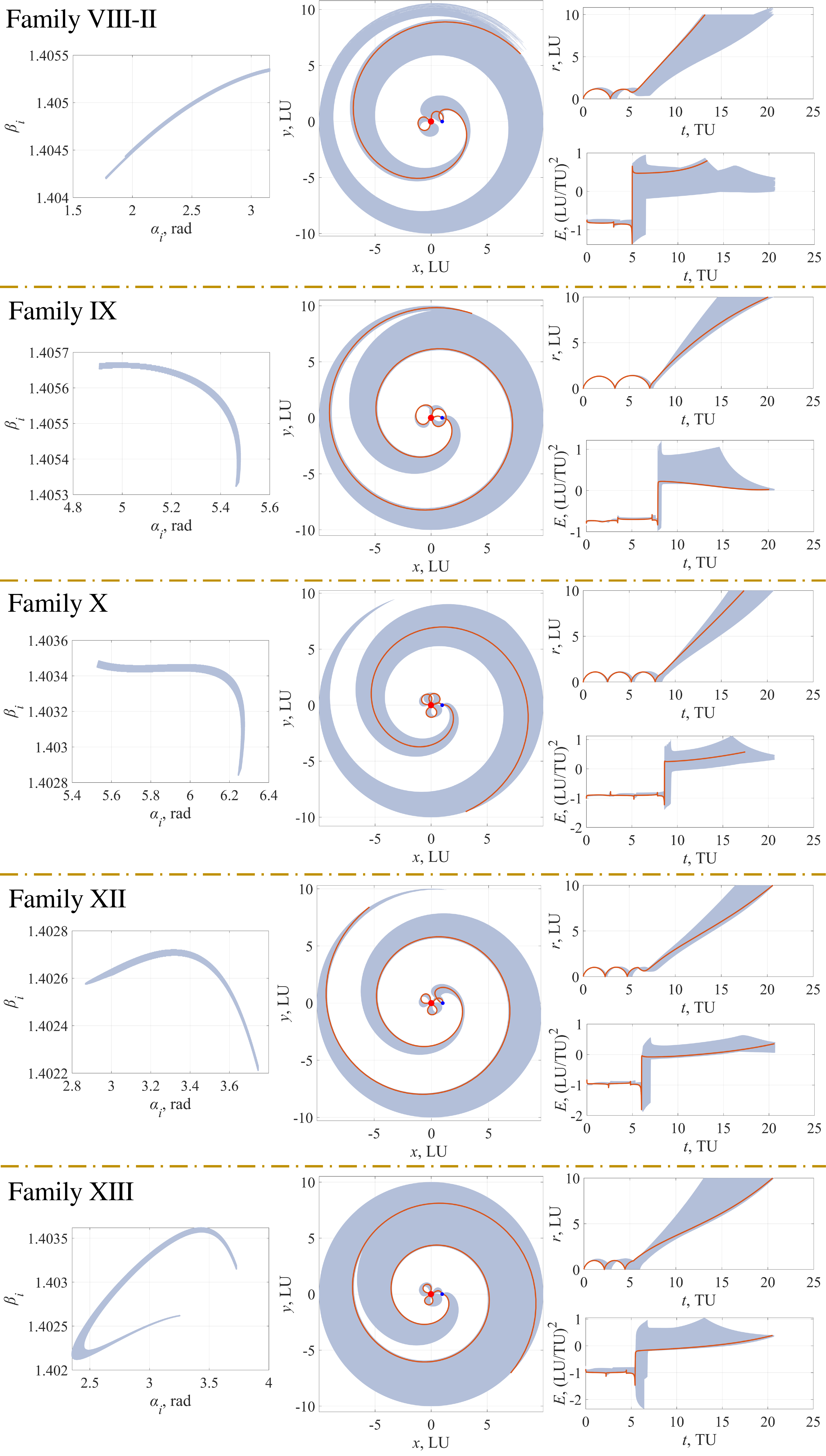}
\caption{The identified escape families in Sun-Earth/Moon PBCR4BP with $\theta_{\text{S}i}=0\text{ }\deg$ (Families VIII-II--XIII).}
\label{fig_4_family_0_deg_3}
\end{figure}

\begin{figure}[H]
\centering
\includegraphics[width=0.75\textwidth]{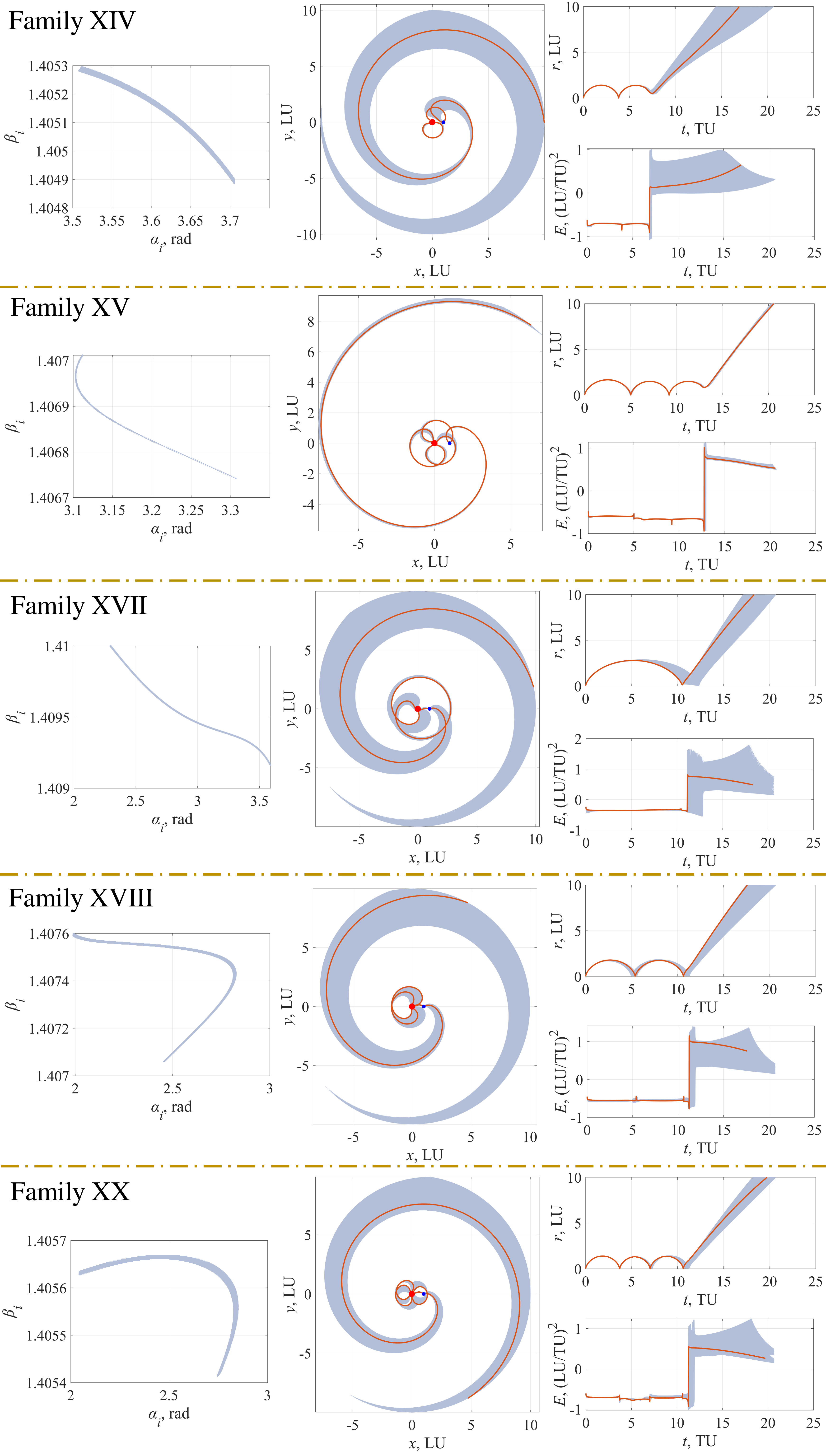}
\caption{The identified escape families in Sun-Earth/Moon PBCR4BP with $\theta_{\text{S}i}=0\text{ }\deg$ (Families XIV--XX).}
\label{fig_4_family_0_deg_4}
\end{figure}

\begin{figure}[H]
\centering
\includegraphics[width=0.75\textwidth]{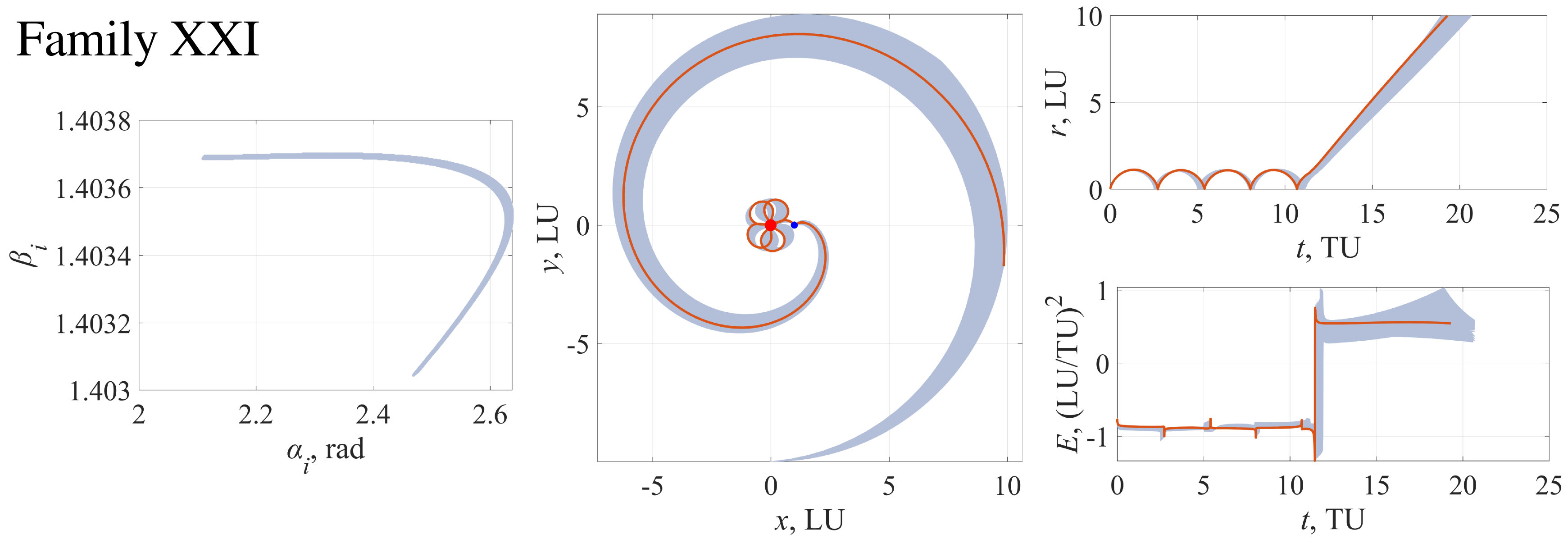}
\caption{The identified escape families in Sun-Earth/Moon PBCR4BP with $\theta_{\text{S}i}=0\text{ }\deg$ (Family XXI).}
\label{fig_4_family_0_deg_5}
\end{figure}

\begin{figure}[H]
\centering
\includegraphics[width=0.75\textwidth]{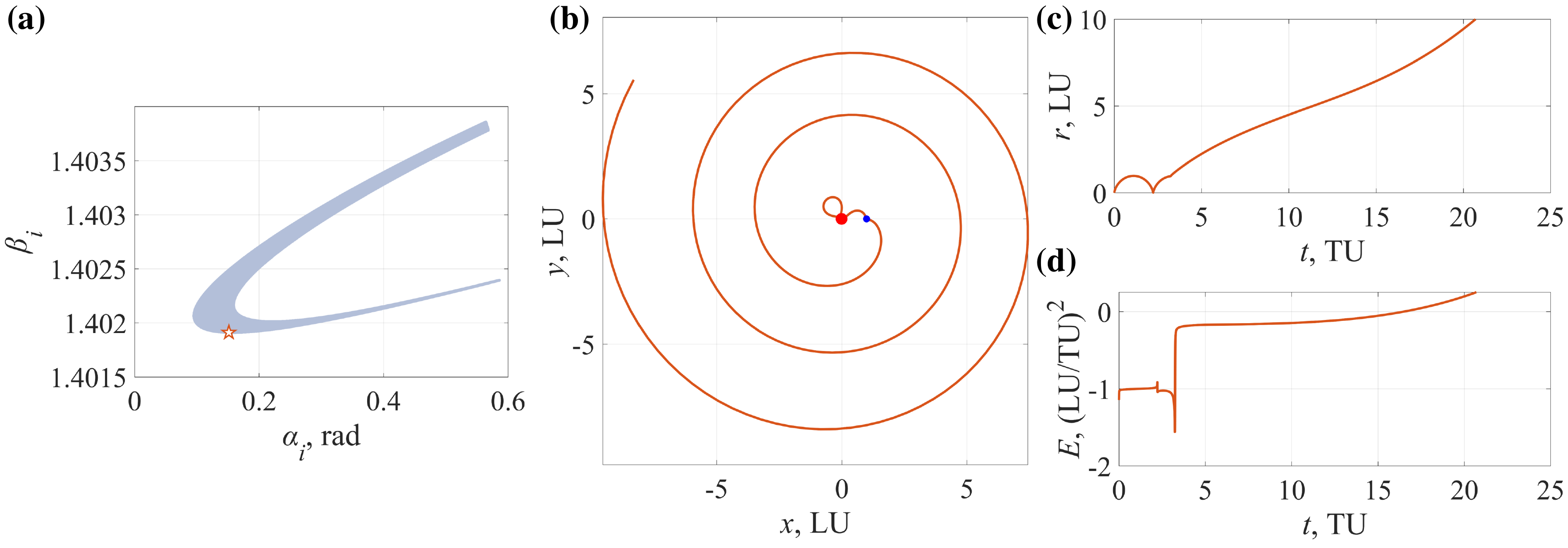}
\caption{Escape trajectory with the minimum $\Delta v_i$ and the corresponding information in the Sun-Earth/Moon PBCR4BP ($\theta_{\text{S}i}=0\text{ }\deg$). (a) Initial states; (b) Trajectory; (c) Time history of $r$; (d) Time history of $E$.}
\label{fig_4_0_deg_min_V}
\end{figure}

We continue to investigate escape trajectories in the Sun-Earth/Moon PBCR4BP. The first case is $\theta_{\text{S}i}=0\text{ }\deg$. Unlike the identified escape families in the Earth-Moon PCR3BP being investigated completely (without noise points), we exclude noise points and some families with few points, and present the representative escape families in the Sun-Earth/Moon PBCR4BP. When $\theta_{\text{S}i}=0\text{ }\deg$, we select 21 escape families, as shown in Figs. \ref{fig_4_family_0_deg_1}-\ref{fig_4_family_0_deg_5}. When labeling these 21 families, we focus on the link between these families and those existing in the Earth-Moon PCR3BP. We consider that the families in the Sun-Earth/Moon PBCR4BP ($\theta_{\text{S}i}=0\text{ }\deg$) are the same families with those in the Earth-Moon PCR3BP if the families share the similar $\left(\alpha_i,\text{ }\beta_i\right)$ distributions and trajectory pattern. Meanwhile, Family $N$-II denotes a continuation of Family $N$, where $N$ denotes the label of families. The distributions in the $\left(\alpha_i,\text{ }\beta_i\right)$ maps of Families $N$ and $N$-II may be interrupted by the constrained TOF and $\beta_i$. Compared to the escape families existing in the Earth-Moon PCR3BP, the new families generated in the Sun-Earth/Moon PBCR4BP ($\theta_{\text{S}i}=0\text{ }\deg$) are Families XX and XXI. The type of LGA of these two families are both prograde LGA. Meanwhile, Families XI, XVI, and XIX are not found for all the obtained escape trajectories in the Sun-Earth/Moon PBCR4BP ($\theta_{\text{S}i}=0\text{ }\deg$). For the time history of $E$, after the LGA, an obvious increase/decrease in the value of $E$ can be observed. This is because the effects of the solar gravity perturbation become obvious when $r$ increases. The corresponding transfer characteristics are summarized in Table \ref{tab4}. For this case, the escape trajectory with the minimum TOF belongs to Family I, whose TOF is 27 days, while escape trajectory with the minimum $\Delta v_i$ belongs to Family III, whose $\Delta v_i$ is 3.132461 km/s. The escape trajectory with the minimum $\Delta v_i$ and the corresponding information are presented in Fig. \ref{fig_4_0_deg_min_V}. Linking with the escape trajectory with the minimum $\Delta v_i$ in the Earth-Moon PCR3BP, it is found that for the considered maximum TOF, the solar gravity perturbation may not play an important role in the reduction of $\Delta v_i$. Subsequently, the case $\theta_{\text{S}i}=90\text{ }\deg$ is investigated.

\begin{table}[h]
\centering
\renewcommand{\arraystretch}{1.5}
\caption{Characteristics of identified escape families in the Sun-Earth/Moon PBCR4BP ($\theta_{\text{S}i}=0\text{ }\deg$).}\label{tab4}%
\begin{tabular}{@{}llllll@{}}
\hline
Family & Number & $\min \Delta v_i$, km/s & $\max \Delta v_i$, km/s & $\min\text{TOF}$, Day  & $\max\text{TOF}$, Day \\
\hline
I        & 475599 & 3.134815 & 3.195498 & 27 & 90 \\
II       & 244994 & 3.137262 & 3.182264 & 46 & 90 \\
III      & 109521 & 3.132461 & 3.147628 & 48 & 90 \\
IV       & 58807  & 3.163933 & 3.175171 & 48 & 90 \\
V        & 33782  & 3.167970 & 3.172101 & 65 & 90 \\
VI       & 18870  & 3.189356 & 3.195498 & 67 & 90 \\
VI-II   & 13404  & 3.193144 & 3.195498 & 58 & 90 \\
VII      & 18448  & 3.185927 & 3.190962 & 60 & 90 \\
VII-II  & 2134   & 3.192848 & 3.195498 & 71 & 90 \\
VIII     & 10657  & 3.135828 & 3.137480 & 60 & 90 \\
VIII-II & 40564  & 3.150309 & 3.159288 & 56 & 90 \\
IX       & 13194  & 3.159054 & 3.161750 & 63 & 90 \\
X        & 39077  & 3.139709 & 3.144713 & 69 & 90 \\
XII      & 22145  & 3.134799 & 3.138758 & 72 & 90 \\
XIII     & 134811 & 3.134098 & 3.145757 & 57 & 90 \\
XIV      & 3987   & 3.155640 & 3.158835 & 62 & 90 \\
XV       & 248    & 3.170105 & 3.172210 & 88 & 90 \\
XVII     & 4980   & 3.188967 & 3.195498 & 74 & 90 \\
XVIII    & 9641   & 3.172584 & 3.176777 & 76 & 90 \\
XX       & 12403  & 3.159755 & 3.161735 & 79 & 90 \\
XXI      & 12402  & 3.141284 & 3.146396 & 82 & 90 \\
\hline
\end{tabular}
\end{table}

\subsection{Escape Families in The Sun-Earth/Moon PBCR4BP ($\theta_{\text{S}i}=90\text{ }\deg$)}\label{subsec4.3}
\begin{figure}[H]
\centering
\includegraphics[width=0.75\textwidth]{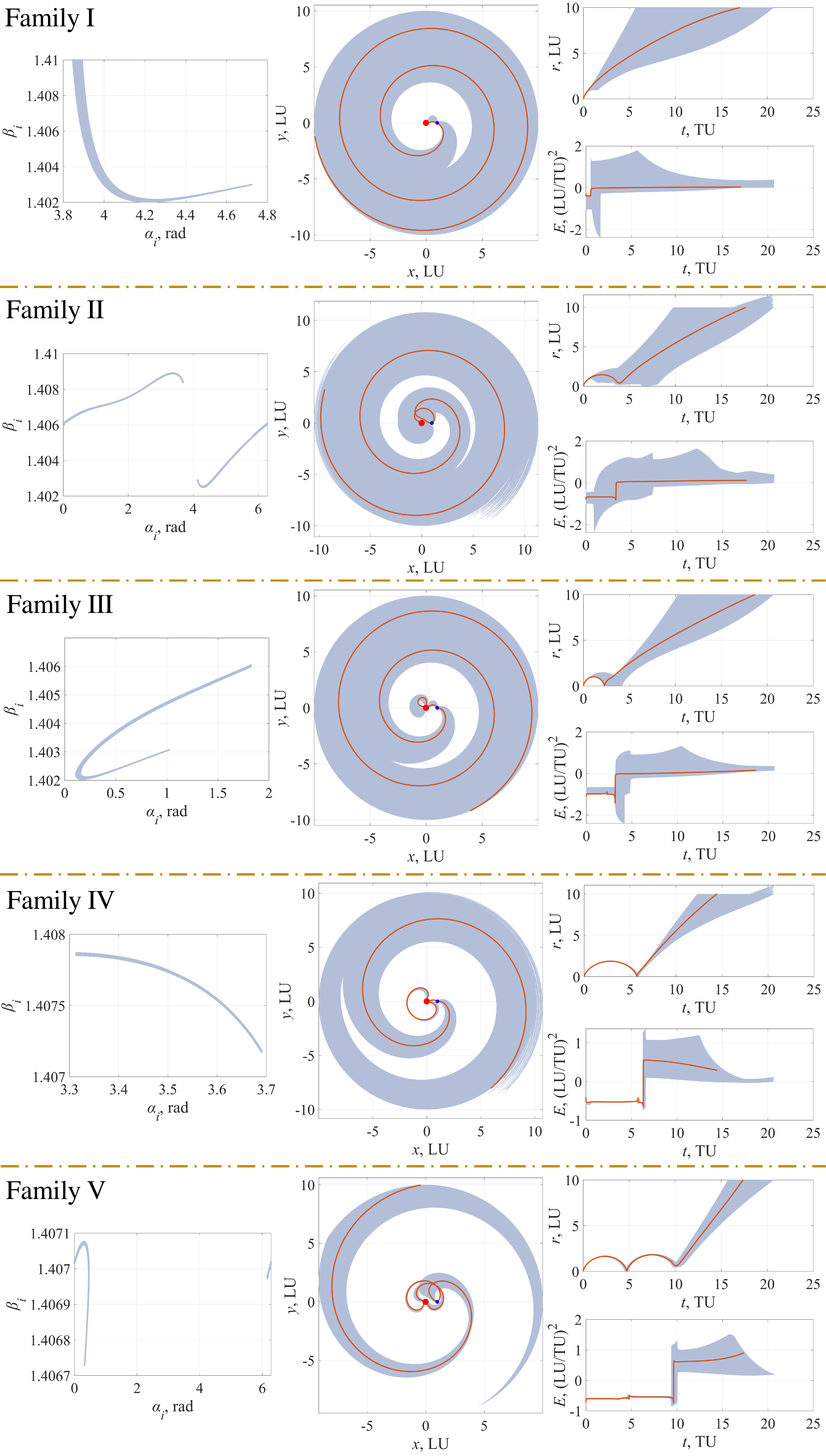}
\caption{The identified escape families in Sun-Earth/Moon PBCR4BP with $\theta_{\text{S}i}=90\text{ }\deg$ (Families I--V).}
\label{fig_4_family_90_deg_1}
\end{figure}

\begin{figure}[H]
\centering
\includegraphics[width=0.75\textwidth]{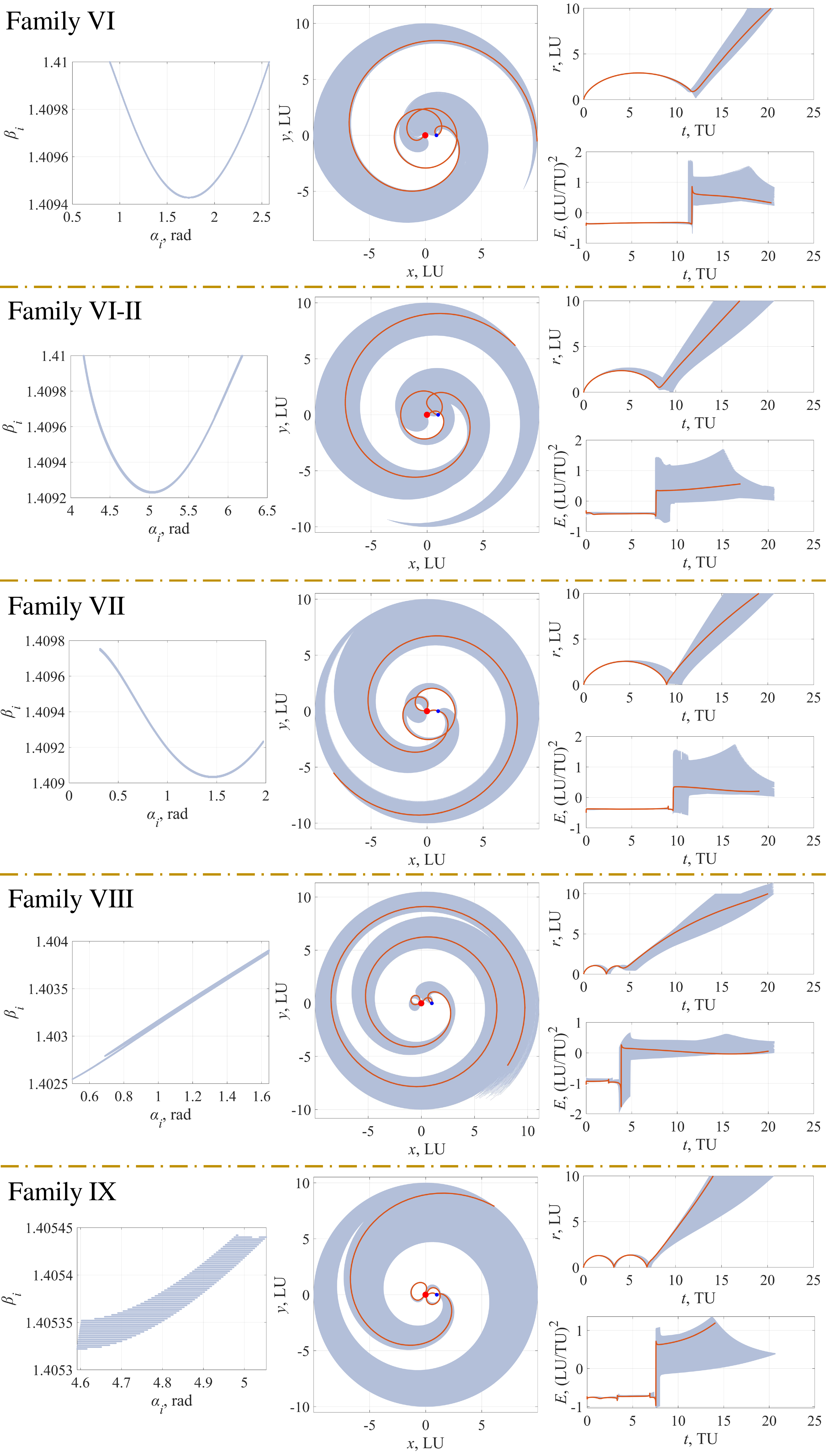}
\caption{The identified escape families in Sun-Earth/Moon PBCR4BP with $\theta_{\text{S}i}=90\text{ }\deg$ (Families VI--IX).}
\label{fig_4_family_90_deg_2}
\end{figure}

\begin{figure}[H]
\centering
\includegraphics[width=0.75\textwidth]{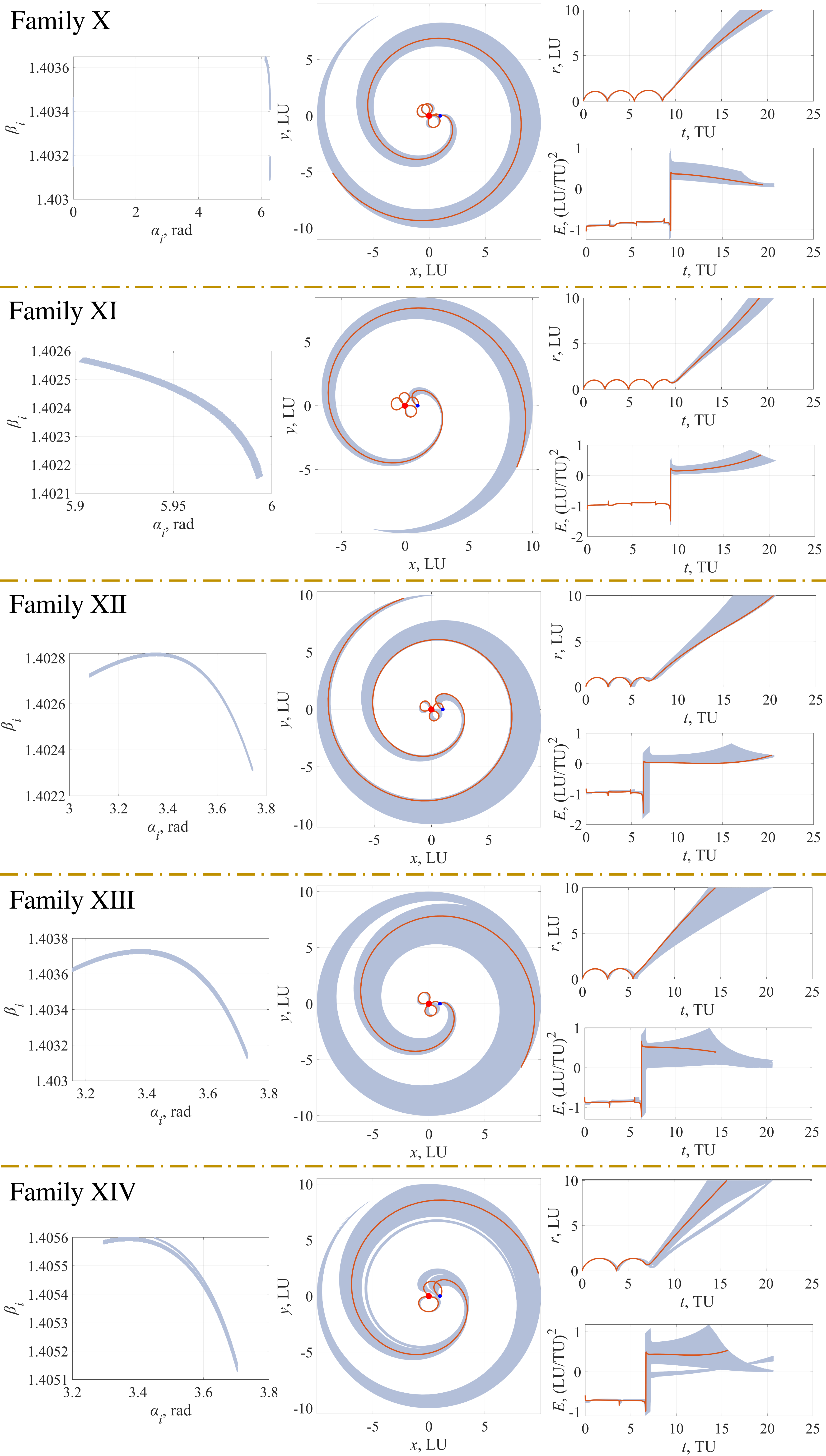}
\caption{The identified escape families in Sun-Earth/Moon PBCR4BP with $\theta_{\text{S}i}=90\text{ }\deg$ (Families X--XVI).}
\label{fig_4_family_90_deg_3}
\end{figure}

\begin{figure}[H]
\centering
\includegraphics[width=0.75\textwidth]{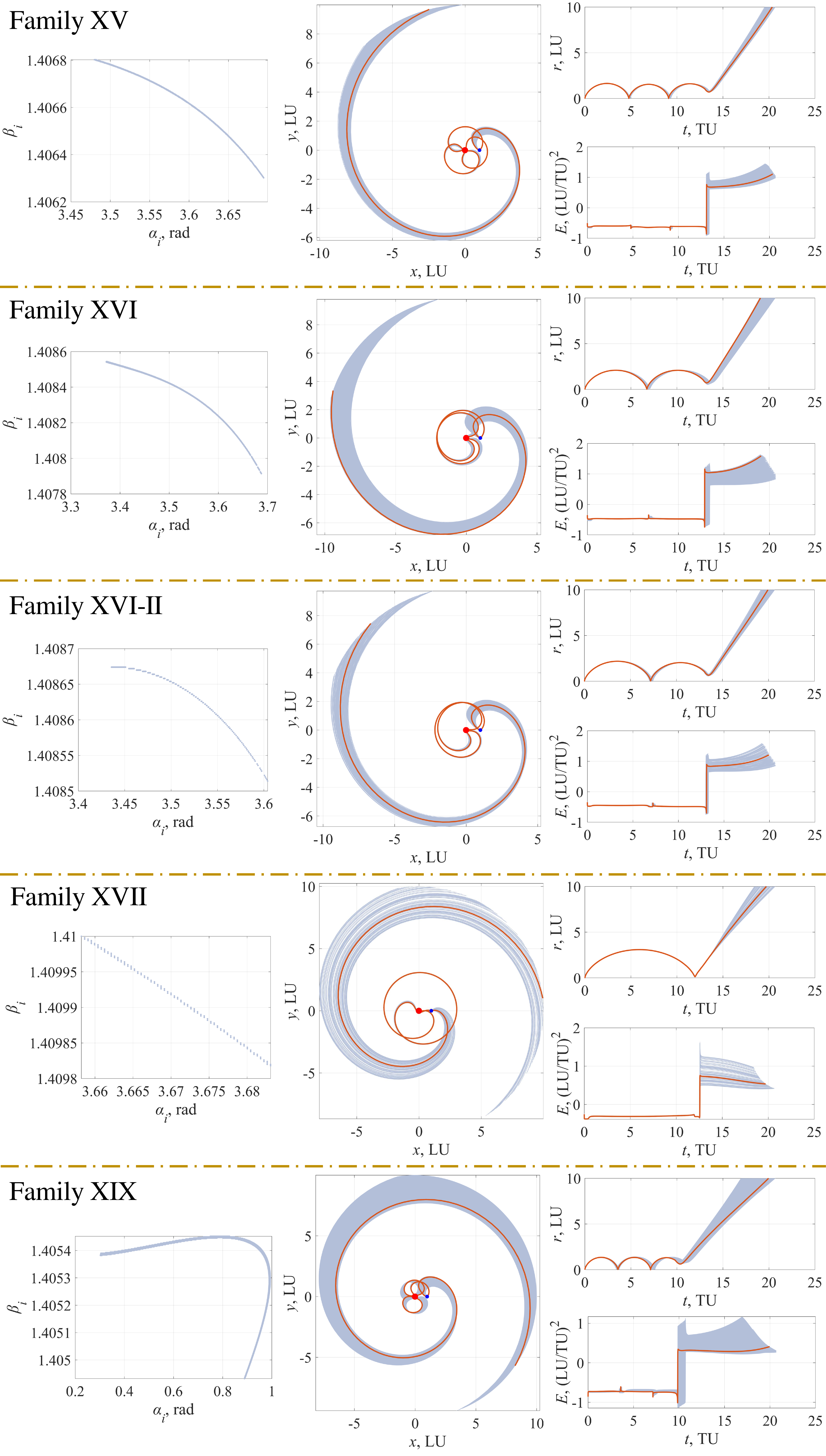}
\caption{The identified escape families in Sun-Earth/Moon PBCR4BP with $\theta_{\text{S}i}=90\text{ }\deg$ (Families XV--XIX).}
\label{fig_4_family_90_deg_4}
\end{figure}

\begin{figure}[H]
\centering
\includegraphics[width=0.75\textwidth]{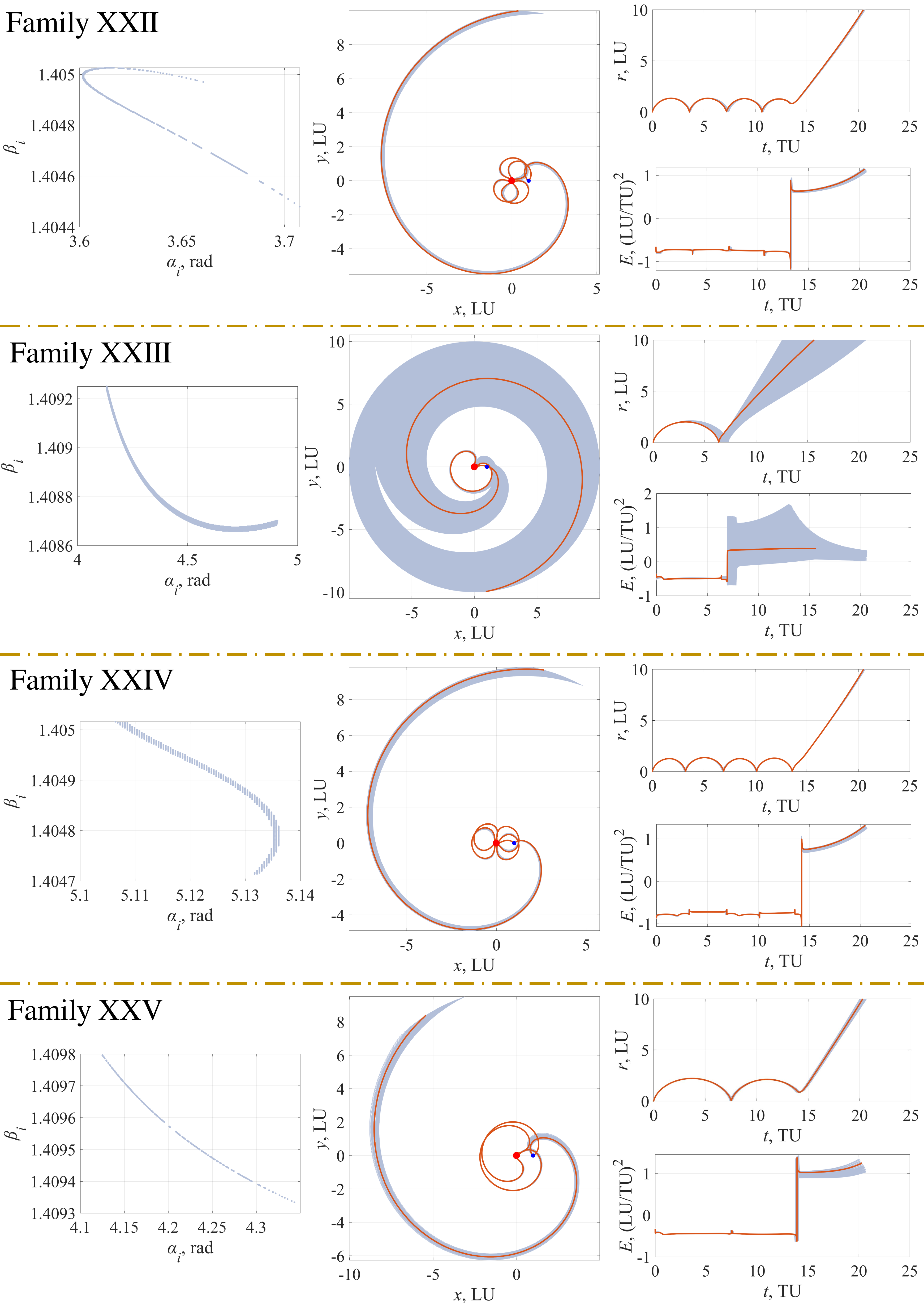}
\caption{The identified escape families in Sun-Earth/Moon PBCR4BP with $\theta_{\text{S}i}=90\text{ }\deg$ (Families XXII--XXV).}
\label{fig_4_family_90_deg_5}
\end{figure}

\begin{figure}[H]
\centering
\includegraphics[width=0.75\textwidth]{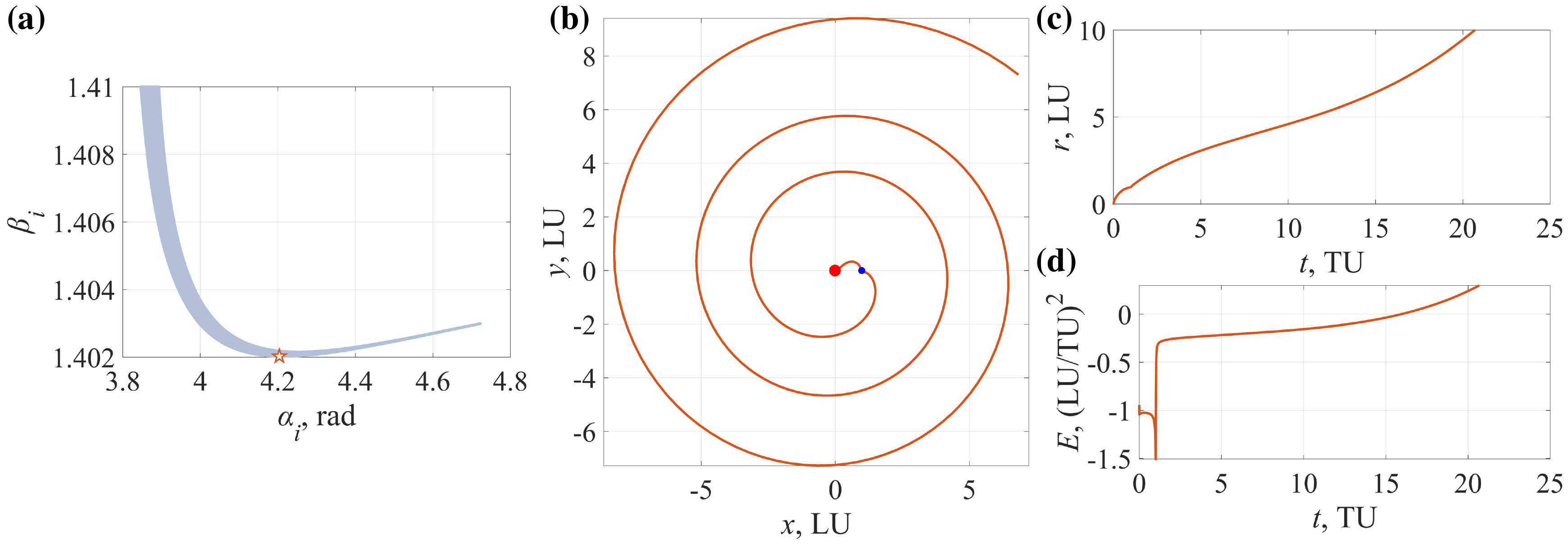}
\caption{Escape trajectory with the minimum $\Delta v_i$ and the corresponding information in the Sun-Earth/Moon PBCR4BP ($\theta_{\text{S}i}=90\text{ }\deg$). (a) Initial states; (b) Trajectory; (c) Time history of $r$; (d) Time history of $E$.}
\label{fig_4_90_deg_min_V}
\end{figure}

When $\theta_{\text{S}i}=90\text{ }\deg$, we select 24 escape families, as shown in Figs. \ref{fig_4_family_90_deg_1}-\ref{fig_4_family_90_deg_5}. Compared to the cases in the Earth-Moon PCR3BP, the new generated families involve Families XXII, XXIII, XXIV, and XXV, while Family XVIII is not observed. Among these new generated families, escape families with the prograde LGA are Families XXIII and XXIV; while escape families with the retrograde LGA are Families XXII and XXV. The corresponding transfer characteristics are summarized in Table \ref{tab5}. For this case, the escape trajectory with the minimum TOF belongs to Family I, whose TOF is 25 days, while escape trajectory with the minimum $\Delta v_i$ also belongs to Family I, whose $\Delta v_i$ is 3.133349 km/s. The escape trajectory with the minimum $\Delta v_i$ and the corresponding information are presented in Fig. \ref{fig_4_90_deg_min_V}. Compared to the escape trajectory with the minimum $\Delta v_i$ in the Earth-Moon PCR3BP, it confirms that the solar gravity perturbation may not play an important role in reducing $\Delta v_i$. 

\begin{table}[htb!]
\centering
\renewcommand{\arraystretch}{1.5}
\caption{Characteristics of identified escape families in the Sun-Earth/Moon PBCR4BP ($\theta_{\text{S}i}=90\text{ }\deg$).}\label{tab5}%
\begin{tabular}{@{}llllll@{}}
\hline
Family & Number & $\min \Delta v_i$, km/s & $\max \Delta v_i$, km/s & $\min\text{TOF}$, Day  & $\max\text{TOF}$, Day \\
\hline
I        & 437203 & 3.133349 & 3.195498 & 25 & 90 \\
II       & 262830 & 3.136950 & 3.187034 & 42 & 90 \\
III      & 185132 & 3.133443 & 3.164681 & 45 & 90 \\
IV       & 6625   & 3.173457 & 3.178897 & 54 & 90 \\
V        & 8381   & 3.169996 & 3.172709 & 68 & 90 \\
VI       & 4962   & 3.191024 & 3.195498 & 77 & 90 \\
VI-II   & 23426  & 3.189481 & 3.195498 & 61 & 90 \\
VII      & 12698  & 3.187953 & 3.193581 & 66 & 90 \\
VIII     & 34737  & 3.137402 & 3.148002 & 62 & 90 \\
IX       & 15100  & 3.159038 & 3.159973 & 59 & 90 \\
X        & 6296   & 3.141627 & 3.145991 & 74 & 90 \\
XI       & 2897   & 3.134331 & 3.137620 & 78 & 90 \\
XII      & 7852   & 3.135563 & 3.139538 & 69 & 90 \\
XIII     & 15688  & 3.141938 & 3.146677 & 59 & 90 \\
XIV      & 6944   & 3.157542 & 3.161189 & 59 & 90 \\
XV       & 639    & 3.166676 & 3.170573 & 85 & 90 \\
XVI      & 944    & 3.179240 & 3.184150 & 83 & 90 \\
XVI-II  & 207    & 3.183916 & 3.185163 & 83 & 90 \\
XVII     & 125    & 3.194079 & 3.195498 & 80 & 90 \\
XIX      & 7566   & 3.155999 & 3.160051 & 74 & 90 \\
XXII     & 343    & 3.152476 & 3.156731 & 89 & 90 \\
XXIII    & 15214  & 3.185039 & 3.189653 & 54 & 90 \\
XXIV     & 538    & 3.154299 & 3.156653 & 89 & 90 \\
XXV      & 163    & 3.190307 & 3.193908 & 87 & 90 \\
\hline
\end{tabular}
\end{table}

\subsection{Effects of The Solar Gravity Perturbation}\label{subsec4.4}
In the above discussion, we perform a systematic analysis on escape families in the Earth-Moon PCR3BP and the Sun-Earth/Moon PBCR4BP. In this subsection, we summarize the effects of the solar gravity perturbation for the considered cases in this paper as follows:
\begin{enumerate}[label=(\arabic*)]
\item \textit{Number of Escape Trajectories}: According to Table \ref{tab1}, the introduction of the solar gravity perturbation cause an increase of the number of escape trajectories for all three cases, i.e., escape trajectories with one LGA (797197 vs. 1280411 ($\theta_{\text{S}i}=0\text{ }\deg$) and 797197 vs. 1058458 ($\theta_{\text{S}i}=90\text{ }\deg$)), two LGAs (68151 vs. 75521 ($\theta_{\text{S}i}=0\text{ }\deg$) and 68151 vs. 85963 ($\theta_{\text{S}i}=90\text{ }\deg$)), and three LGAs ($\theta_{\text{S}i}=90\text{ }\deg$)), two LGAs (622 vs. 1075 ($\theta_{\text{S}i}=0\text{ }\deg$) and 622 vs. 954 ($\theta_{\text{S}i}=90\text{ }\deg$)). Generally, the solar gravity perturbation facilitates the escape from the Earth-Moon system. In particular, for three main families (Family I, Family II, and Families III+IV) of escape trajectories with one LGA, the solar gravity perturbation causes an increase of number of escape trajectories for Families I and II (262375 vs. 475599 ($\theta_{\text{S}i}=90\text{ }\deg$) and 262375 vs. 437203 ($\theta_{\text{S}i}=90\text{ }\deg$) for Family I, 174527 vs. 244994 ($\theta_{\text{S}i}=0\text{ }\deg$) and 174527 vs. 262830 ($\theta_{\text{S}i}=90\text{ }\deg$) for Family II). Considering Families III+IV, when $\theta_{\text{S}i}=0\text{ }\deg$, the number of escape trajectories (168328) is comparable that in the Earth-Moon PCR3BP (171898); when $\theta_{\text{S}i}=90\text{ }\deg$, an obvious increase in the number of escape trajectories (191757) can be observed. 
\item \textit{Escape Families}: The solar gravity perturbation causes the emergence and disappearance of escape families. New generated escape families compared to the Earth-Moon PCR3BP involves Families XX and XXI ($\theta_{\text{S}i}=0\text{ }\deg$), and Families XXII, XXIII, XXIV, and XXV ($\theta_{\text{S}i}=90\text{ }\deg$). Disappeared families involve Families XI, XVI, and XIX ($\theta_{\text{S}i}=0\text{ }\deg$), and Family XVIII ($\theta_{\text{S}i}=90\text{ }\deg$).
\item \textit{Variation in} $E$: Unlike the value of $E$ remaining constant approximately after the LGA in the Earth-Moon PCR3BP, when considering the solar gravity perturbation, an obvious increase/decrease of $E$ can be observed after the LGA. A typical example can be found in the time history of $E$ of Families XVI and XVII when $\theta_{\text{S}i}=90\text{ }\deg$.
\item \textit{Transfer Characteristics}: Comparing the escape trajectories with the minimum $\Delta v_i$ for the considered three cases (i.e., the Earth-Moon PCR3BP, the Sun-Earth/Moon PBCR4BP ($\theta_{\text{S}i}=0\text{ }\deg$), and the Sun-Earth/Moon PBCR4BP ($\theta_{\text{S}i}=90\text{ }\deg$)), the solar gravity perturbation shows a limited role in effectively reducing $\Delta v_i$.
\end{enumerate}

Based on the above discussion, the effects of the solar gravity perturbation are mainly revealed in the \textit{Number of Escape Trajectories} and \textit{Escape Families}. In particular, even considering the solar gravity perturbation, three main families (Family I, Family II, and Families III+IV) still exist. The solar gravity perturbation shows limited effects on escape trajectories with one LGA possibly because these trajectories escape from the Earth-Moon system rapidly after the LGA and the solar gravity perturbation has no sufficient time to affect trajectories effectively. Therefore, for practical mission scenarios that typically require the design of one or a limited number of feasible escape trajectories, the above three main families can be applied and the Earth–Moon PCR3BP can be adopted in the preliminary design stage, instead of the Sun–Earth/Moon PBCR4BP, to achieve a balance between model complexity and overall design performance in terms of transfer characteristics (e.g., $\Delta v_i$). Notably, it should be clarified that under the same initial states, escape can take place within 90 days in the Earth-Moon PCR3BP but may not take place in the Sun-Earth/Moon PBCR4BP (examples of escape trajectories belonging to Family I are presented in Appendix). Therefore, although the Earth-Moon PCR3BP can be adopted in the preliminary design stage, the obtained escape trajectories should be further checked in or continued to the Sun–Earth/Moon PBCR4BP or other high-fidelity models \cite{campana2025ephemeris,sanaga2025leveraging}. Also, once the escape trajectories are constructed, the design of launch vehicle \cite{jiang2026adaptive} and the optimization of ascent trajectory \cite{miao2021successive} should be performed to satisfy practical mission requirements.

\section{Conclusion}\label{sec5}
This paper is devoted to proposing an effective identification method of escape families and performing a systematic exploration of escape trajectories in the Earth-Moon planar circular restricted three-body problem (PCR3BP) and the Sun-Earth/Moon planar bicircular restricted four-body problem (PBCR4BP). Firstly, the solution space of initial states of escape trajectories in the Earth-Moon PCR3BP and the Sun-Earth/Moon PBCR4BP with several fixed initial solar phase angles are generated. A dynamical analysis is performed to pre-filter the escape trajectories to be investigated. As a result, escape trajectories with one lunar gravity assist are treated as the focus. Then, an identification method of escape families based on initial state distributions is proposed using the density-based spatial clustering of applications with noise (DBSCAN) method. The effectiveness of the proposed method is verified by the identification results. Once the escape families are identified, the corresponding distribution of initial states, trajectories, and transfer characteristics are analyzed. The comparison between escape families in the Earth-Moon PCR3BP and the Sun-Earth/Moon PBCR4BP is further performed to reveal the effects of the solar gravity perturbation. It is found that the solar gravity perturbation generally facilitates the escape from the Earth-Moon system and causes an obvious variation in the value of generalized energy after the LGA, but plays a limited role in effectively reducing the Earth injection impulse. By analyzing the effects of the solar gravity perturbation on escape trajectories, it is suggested to adopt the Earth-Moon PCR3BP to construct escape trajectories in the preliminary design stage for practical mission application, thereby providing a justification for the model selection based on the quantitative analysis. This paper establishes an analysis methodology of escape trajectories from a perspective of escape families in the Earth-Moon PCR3BP and the Sun-Earth/Moon PBCR4BP, providing further insights into the selection of model and initial states in the trajectory design. 

\section*{Appendix}
\begin{figure}[H]
\centering
\includegraphics[width=0.75\textwidth]{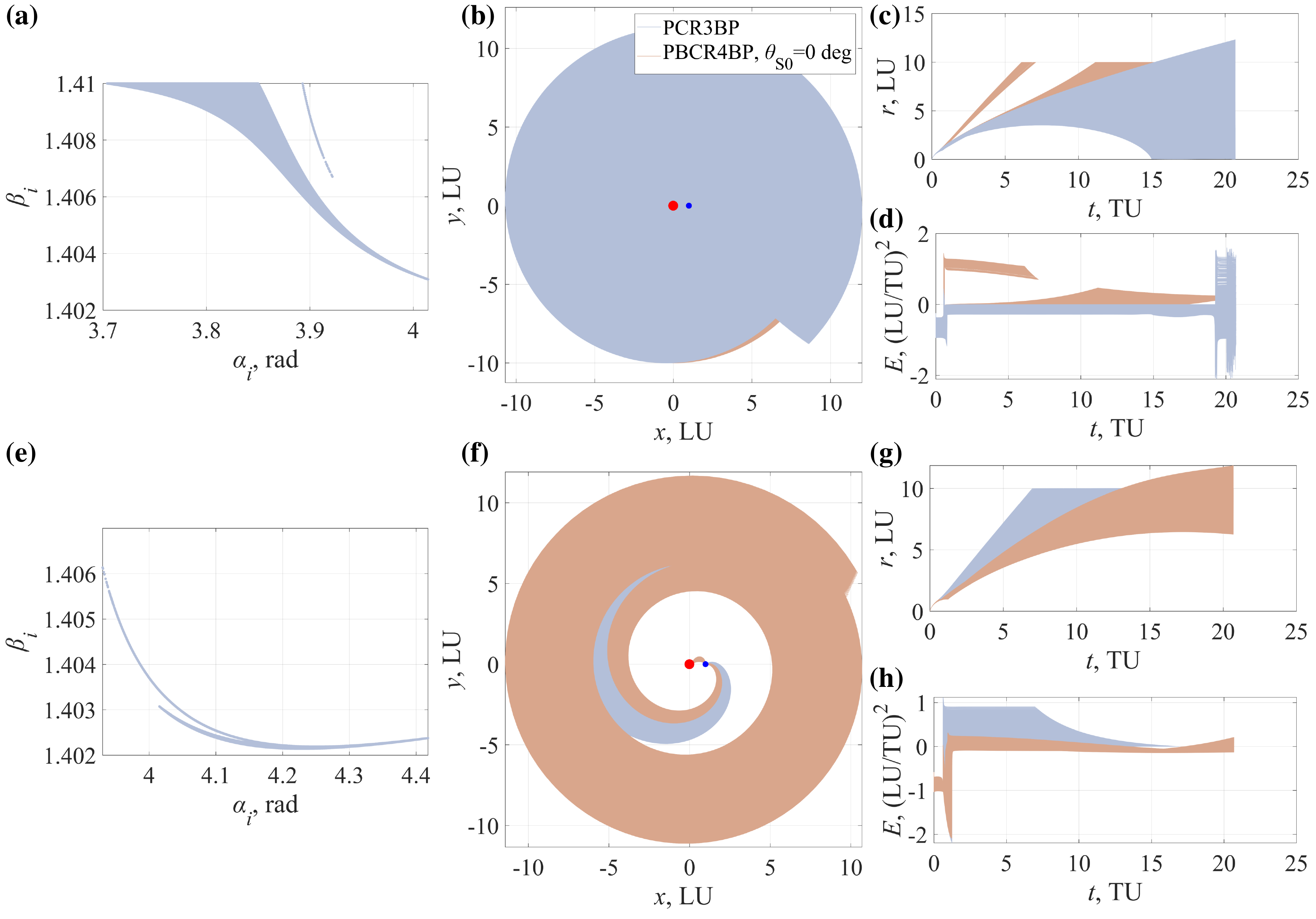}
\caption{Difference between escape trajectories in the Earth-Moon PCR3BP and the Sun-Earth/Moon PBCR4BP (Family I).}
\label{fig_3_4_difference}
\end{figure}

The difference of escape trajectories belonging to Family I in the Earth-Moon PCR3BP and the Sun-Earth/Moon PBCR4BP is presented in Fig. \ref{fig_3_4_difference}, where subfigures. (a)-(d) are corresponding to the trajectories escaping in the Earth-Moon PCR3BP do not escape in the Sun-Earth/Moon PBCR4BP($\theta_{\text{S}i}=0\text{ }\deg$), and subfigures. (f)-(h) are corresponding to the trajectories escaping in the Sun-Earth/Moon PBCR4BP($\theta_{\text{S}i}=0\text{ }\deg$) do not escape in the Earth-Moon PCR3BP. For these two cases, the reasons why the trajectories do not escape can be summarized as follows: (i) trajectories impact the Earth/Moon; (ii) trajectories satisfying $E>0$ do not satisfy $r>10\text{ LU}$; (iii) trajectories satisfying $r>10\text{ LU}$ do not satisfy $E>0$.

\section*{Funding  Sources}

The authors acknowledge the financial support from the National Natural Science Foundation of China (Grant No. 12525204), the National Natural Science Foundation of China (Grant No. 12372044), the National Natural Science Foundation of China (No. U23B6002), the National Natural Science Foundation of China (Grant No. 12302058), and the Young Elite Scientists Sponsorship Program by CAST (Grant No. 2023QNRC001).

\bibliography{sample}

\end{document}